\documentclass[twocolumn,amsmath,amssymb,floatfix,aps,prb,shownopacs,footinbib,superscriptaddress]{revtex4-2}
\usepackage{amsmath,amssymb,amsfonts,bm,graphicx,url,epsfig,color,epstopdf}
\usepackage[british]{babel}
\setcitestyle{super} 
\usepackage{soul} 
\usepackage{siunitx} 
\usepackage[dvipsnames]{xcolor} 
\begin{document}
\title{Signature of anyonic statistics in the integer quantum Hall regime}

\author{P. Glidic}
\affiliation{Universit\'e Paris-Saclay, CNRS, Centre de Nanosciences et de Nanotechnologies, 91120, Palaiseau, France}
\author{I. Petkovic}
\email[e-mail: ]{ivana.petkovic@c2n.upsaclay.fr}
\affiliation{Universit\'e Paris-Saclay, CNRS, Centre de Nanosciences et de Nanotechnologies, 91120, Palaiseau, France}
\author{C. Piquard}
\affiliation{Universit\'e Paris-Saclay, CNRS, Centre de Nanosciences et de Nanotechnologies, 91120, Palaiseau, France}
\author{A. Aassime}
\affiliation{Universit\'e Paris-Saclay, CNRS, Centre de Nanosciences et de Nanotechnologies, 91120, Palaiseau, France}
\author{A. Cavanna}
\affiliation{Universit\'e Paris-Saclay, CNRS, Centre de Nanosciences et de Nanotechnologies, 91120, Palaiseau, France}
\author{Y. Jin}
\affiliation{Universit\'e Paris-Saclay, CNRS, Centre de Nanosciences et de Nanotechnologies, 91120, Palaiseau, France}
\author{U. Gennser}
\affiliation{Universit\'e Paris-Saclay, CNRS, Centre de Nanosciences et de Nanotechnologies, 91120, Palaiseau, France}
\author{C. Mora}
\affiliation{Universit\'e Paris Cit\'e, CNRS, Laboratoire Mat\'eriaux et Ph\'enom\`enes Quantiques, F-75013 Paris, France}
\author{D. Kovrizhin}
\affiliation{CY Cergy Paris Universit\'e, CNRS, Laboratoire de Physique Th\'eorique et Mod\'elisation, Cergy-Pontoise, F-95302, France}
\author{A. Anthore}
\affiliation{Universit\'e Paris-Saclay, CNRS, Centre de Nanosciences et de Nanotechnologies, 91120, Palaiseau, France}
\affiliation{Université Paris Cité, CNRS, Centre de Nanosciences et de Nanotechnologies, F-91120, Palaiseau, France}
\author{F. Pierre}
\email[e-mail: ]{frederic.pierre@cnrs.fr}
\affiliation{Universit\'e Paris-Saclay, CNRS, Centre de Nanosciences et de Nanotechnologies, 91120, Palaiseau, France}

\begin{abstract}
Anyons are exotic low-dimensional quasiparticles whose unconventional quantum statistics extends the binary particle division into fermions and bosons.
The fractional quantum Hall regime provides a natural host, with first convincing anyon signatures recently observed through interferometry and cross-correlations of colliding beams. However, the fractional regime is rife with experimental complications, such as an anomalous tunneling density of states, which impede the manipulation of anyons.
Here we show experimentally that the canonical integer quantum Hall regime can provide a robust anyon platform. 
Exploiting the Coulomb interaction between two co-propagating quantum Hall channels, 
an electron injected into one channel splits into two fractional charges behaving as abelian anyons.
Their unconventional statistics is revealed by negative cross-correlations between dilute quasiparticle beams.
Similarly to fractional quantum Hall observations, we show that the negative signal stems from a time-domain braiding process, here involving the incident fractional quasiparticles and spontaneously generated electron-hole pairs.
Beyond the dilute limit, a theoretical understanding is achieved  
via the edge magnetoplasmon description of interacting integer quantum Hall channels.
Our findings establish that, counter-intuitively, the integer quantum Hall regime 
provides a platform of choice for exploring and manipulating quasiparticles with fractional quantum statistics.
\end{abstract}

\maketitle

Integer and fractional quantum Hall effects \cite{girvin_review} are thought of as fundamentally separate. 
The main features of the integer quantum Hall (IQH) states are well described within the single-particle fermionic picture \cite{girvin_review,vonKlitzing_1980,Laughlin_IQHE_1981,Halperin_EdgeStates_1982}. 
In contrast, fractional quantum Hall (FQH) states inherently stem from strong Coulomb interactions, giving rise to anyons - composite quasiparticles which carry a fractional charge and exhibit anyonic exchange statistics  \cite{Leinas_anyons,Feldman_FQHEdge_2021}. 
Abelian anyons acquire a phase upon exchange, whereas non-abelian anyons undergo a deeper  
transformation into different states\cite{Wen_NonAbelian_1991}. 
Demonstrating the exchange statistics of anyons, in particular non-abelian, is a crucial stepping stone towards realizing topological quantum computing  \cite{Nayak_TopoRMP_2008}.

\begin{figure*}[!htb]
\renewcommand{\figurename}{\textbf{Figure}}
\renewcommand{\thefigure}{\textbf{\arabic{figure}}}
	\centering
	\includegraphics[width=17.9cm]{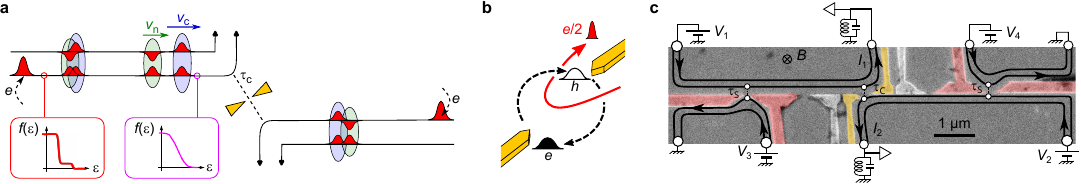}
	\caption{
	\footnotesize
        \textbf{Experimental setup.}
		\textbf{a,} In the presence of two strongly coupled quantum Hall channels at $\nu=2$, tunneling electrons $e$ (individual red wave-packets) progressively split into two pairs (circled).
    The fast `charge' pair (blue background) consists of two co-propagating $e/2$ wave-packets, one in each channel, whereas the slow `neutral' pair (green background) consists of opposite $\pm e/2$ charges.
    The fractionalized $e/2$ charges propagate toward a central QPC (yellow split gates) of transmission $\tau_\mathrm{c}$, used to investigate their quantum statistics from the outgoing current cross-correlations.
    The strong coupling regime and the degree of fractionalization at the level of the central QPC are established separately through the evolution of the electron energy distribution function $f(\varepsilon)$ from a non-equilibrium double step (red inset) to a smoother function (magenta inset).
    \textbf{b,}  Illustration of the time-braiding mechanism, whereby an impinging fractionalized $e/2$ charge (red) braids with an electron-hole pair (black) spontaneously excited at the central QPC. 
	\textbf{c,} E-beam micrograph of the sample. 
 The two copropagating edge channels are drawn as black lines with arrows indicating the chirality.
 The aluminum gates used to form the QPCs by field effect are highlighted in false colors (sources in red, central analyzer in yellow).
 A negative voltage is applied to the non-colored gates to reflect the edge channels at all times.
 Tunneling at the sources is controlled by the applied dc voltages $V_\mathrm{1,2,3,4}$ and through their gate-controlled transmission probability $\tau_\mathrm{s}$.
}	\label{FigSchemSample}
\end{figure*}

Anyonic exchange statistics can be revealed via interferometry, whereby anyons along the edge move around those in the bulk, and acquire a  braiding (double exchange) phase 
\cite{Feldman_FQHEdge_2021, Nakamura_Anyon1s3_2020, Nakamura_Anyon2s5_2023, Kundu_AnyonMZ}.
An alternative probe, not requiring involved heterostructures with built-in screening, is provided by a mixing process at an `analyzer' quantum point contact (QPC) 
\cite{Lee_NegativeNoiseBraiding_2019}. 
If the impinging quasiparticle beams are dilute (Poissonian), the outgoing current cross-correlations carry a signature of anyonic statistics \cite{Lee_NegativeNoiseBraiding_2019,Lee_BunchingOrBraiding_2023,Rosenow_Collider_2016}. 
These dilute beams are created upstream by sources typically realized by voltage-biased QPCs set in the tunneling regime.
The signature of 
anyonic statistics becomes particularly straightforward with two symmetric sources, a configuration often referred to as a `collider' \cite{Rosenow_Collider_2016}.
In this simple case, the cross-correlations for free fermions vanish, whereas a negative signal is considered to be a strong marker of anyonic statistics\cite{Rosenow_Collider_2016, Feldman_FQHEdge_2021, Lee_NonAbelianCollider_2022}.
Such negative current cross-correlation signatures of anyons at filling factors $\nu=1/3$ and $2/5$ have recently been demonstrated  \cite{Bartolomei_Cross_1tiers_2020,Glidic_Collider2023, Lee_BunchingOrBraiding_2023, Ruelle_2s5_2023}. 

However, FQH states present complications which impede 
the analysis and further manipulation of anyons: the edge structure is often undetermined between several alternatives \cite{Heiblum_Edge_Review},
the tunneling density of states generally presents an anomalous voltage dependence \cite{Chang_LL_2003}, and the decoherence along the edge appears to be very strong \cite{Kundu_AnyonMZ,Nakamura_Anyon1s3_2020,Nakamura_Anyon2s5_2023}. 
A promising alternative path is provided by the insight that fractional charges propagating along the edges of IQH states should also behave as  anyons\cite{Wen_Review_1995,Lee_MutualStats_2020,Morel_IntegerCollider_2022}, although they are not topologically protected, unlike fractional bulk quasiparticles. 
Indeed, the exchange phase of two quasiparticles of charge $e^*$ propagating along an integer quantum Hall 
 channel \cite{Lee_MutualStats_2020} is $\pi (e^*/e)^2$. This exchange phase can be linked to a dynamical Aharonov-Bohm effect \cite{Stern_2008}.
In practice, such IQH anyons could be obtained e.g. by driving the edge channel with a narrow voltage pulse,
from the charge fractionalization across a Coulomb island, or by exploiting the intrinsic 
Coulomb coupling between co-propagating edge channels \cite{Safi_InhomoTLL_1995, Berg_FractionalIntEdge_2009, Lee_MutualStats_2020,Morel_IntegerCollider_2022}. 
The present work proposes and implements the latter strategy, and demonstrates the anyonic character of the resulting fractional charges from the emergence of negative cross-correlations.

We focus on the filling factor $\nu=2$, which has two copropagating edge channels and constitutes the most simple, canonical and robust IQH state with interacting channels. 
The edge physics is well described by a chiral Luttinger model involving two one-dimensional channels with a linear dispersion relation and short-range Coulomb interactions \cite{Wen_Chiral_LL,Chang_LL_2003,Idrisov_ColliderTwo_2022,Kovrizhin_Equilibration_IQHE_2011,Kovrizhin_EMP_2012}.
This theory, which has been successful in explaining many experimental findings such as multiple lobes in a Mach-Zehnder interferometer \cite{Neder_MZLobes_2006,Levkivskyi_halfe_2008,Sim_MZ_2008}, spin-charge separation \cite{Hashisaka_SpinChargeSeparation_2017,Bocquillon_SeparationModes_2013}  or noise measurements \cite{Inoue_Fractionalization_2014}, reformulates interacting fermionic edge states as two free edge magnetoplasmon (EMP) modes via bosonization. 
In the limit of weak inter-channel coupling, each EMP mode is localized in one different channel and the system can be mapped back into the free electron picture.
In contrast, at strong coupling the two EMP modes are fully delocalized between the two quantum Hall channels and correspond to a charge mode, with identical charge density fluctuations on both channels, and a neutral mode, with opposite density fluctuations \cite{Wen_halfe_1990,Wen_Review_1995,Berg_FractionalIntEdge_2009}.
Experimentally, typical Al(Ga)As devices at $\nu=2$ often appear to be close to the strong coupling regime \cite{leSueur_ERIQHE_2010,Hashisaka_SpinChargeSeparation_2017,Bocquillon_SeparationModes_2013}.

Here we exploit such an inter-channel distribution of EMPs at strong coupling to split electrons into fractional charges, similarly to the theoretical proposal in [\!\!~\citenum{Berg_FractionalIntEdge_2009}]. We start by injecting electrons into a single edge channel with a voltage-biased QPC.
Then, downstream from the QPC, each injected electron progressively splits into two wave-packets.
Assuming strong coupling, one is solely built upon charge EMPs propagating at velocity $v_\mathrm{c}$, and the other is constructed from neutral EMPs and has a slower velocity $v_\mathrm{n}$.
If we consider separately the quantum Hall channel where the electron is injected, both wave-packets carry a fractional charge of $e/2$, whereas in the other channel they have opposite charges $\pm e/2$ (see Fig.~\ref{FigSchemSample}a)\cite{Berg_FractionalIntEdge_2009,Acciai_MA_2018}.
Such fractional wave-packets propagate non-dispersively. Considered individually, they are predicted to behave as abelian anyons with non-trivial exchange phase \cite{Wen_Review_1995,Feldman_FQHEdge_2021,Lee_MutualStats_2020,Morel_IntegerCollider_2022}. 

To experimentally address the anyon character of fractional charges propagating along integer quantum Hall channels, we measure the current cross-correlations at the output of a `collider' in the stationary regime (see Fig.~\ref{FigSchemSample}a,c). 
As for the fractional quantum Hall version of the device\cite{Rosenow_Collider_2016}, direct anyon collisions are very rare and can be ignored in the relevant dilute beam limit\cite{Lee_NonAbelianCollider_2022,Morel_IntegerCollider_2022}.
The cross-correlation signal stems instead from a braiding in the time-domain between incident anyons and particle-hole pairs spontaneously excited at the QPC\cite{Han_TopoVacuumBubble_2016, Lee_NegativeNoiseBraiding_2019, Lee_MutualStats_2020, Lee_NonAbelianCollider_2022, Morel_IntegerCollider_2022, Lee_BunchingOrBraiding_2023,Iyer_2023,Thamm_2023}, as illustrated in Fig.~\ref{FigSchemSample}b. 
At integer filling factors, the pairs are always formed of fermionic  particles (electrons and holes), whereas in the fractional quantum Hall regime they can consist of anyons.
Therefore the time-braiding considered here takes place between two different types of quasiparticles, of fractional and integer charges.
The braiding (double exchange) phase $2\theta$ acquired in such heterogeneous cases characterizes the so-called mutual quantum statistics.
It is predicted to take the fractional value of $2\theta=\pi$ (compared to $0\,(\mathrm{mod}\,2\pi)$ for the braiding of fermions or bosons).
Note that while the braiding mechanism is equally relevant for a single incident beam of fractional quasiparticles or for two symmetric beams, the latter `collider' setup allows for a qualitative test of the unconventional anyon character from the mere emergence of non-zero cross-correlations at the output\cite{Rosenow_Collider_2016, Lee_NonAbelianCollider_2022, Iyer_2023, Thamm_2023}.

\begin{figure*}[!bth]
\renewcommand{\figurename}{\textbf{Figure}}
\renewcommand{\thefigure}{\textbf{\arabic{figure}}}
	\centering
	\includegraphics[width=17.9cm]{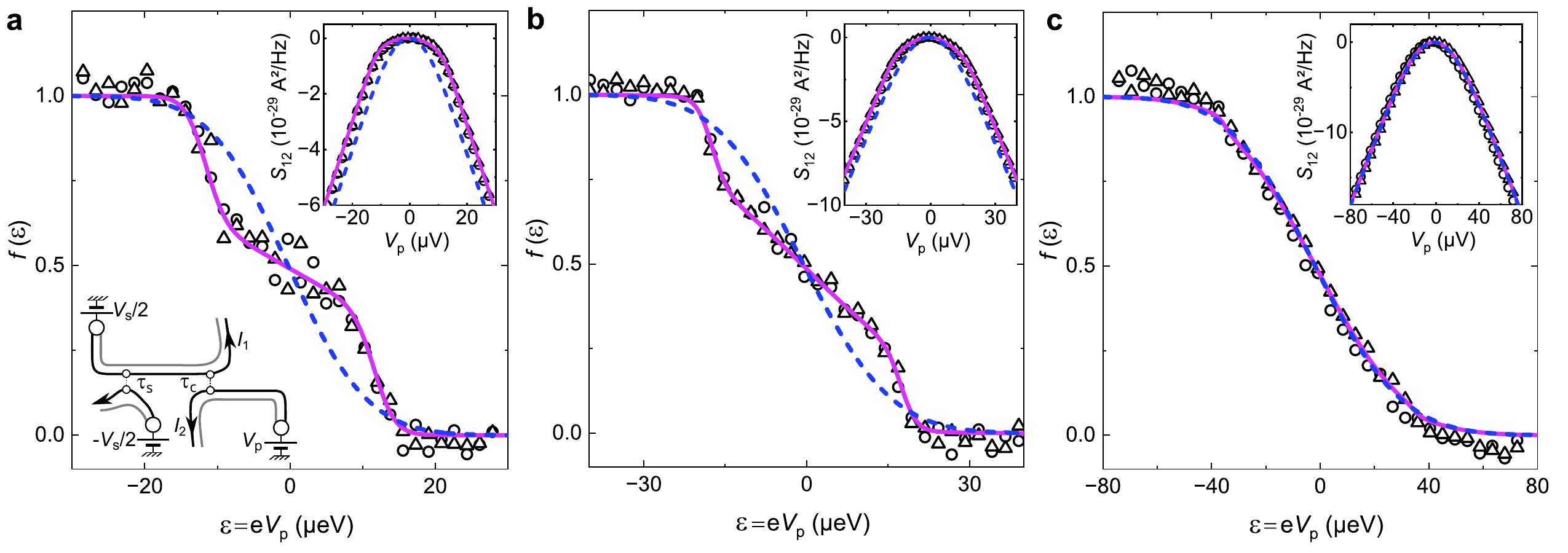}
	\caption{
	\footnotesize
 \textbf{Spectroscopy of the electron energy distribution \boldmath$f(\varepsilon)$.}
The shape of $f(\varepsilon)$ reflects the inter-channel coupling regime and informs on the conditions for a complete charge fractionalization at the central QPC.
One source is voltage biased at $V_\mathrm{s}$, here with $\tau_\mathrm{s}\approx0.5$, and the same probe voltage $V_\mathrm{p}$ is applied across the other one (see schematic in \textbf{a}).
Circles and triangles show data points with the voltage biased source QPC on the left and right side, respectively. 
Purple continuous lines and blue dashed lines represent exact theoretical predictions in the strong coupling regime for a time delay between charge and neutral pairs of $\delta t=$ \SI{64}{ps} and $\infty$, respectively (see Supplementary Information).
  Insets: Cross-correlations $S_{12}$ versus probe voltage $V_\mathrm{p}$.
   Main panels: $f(\varepsilon)$ obtained by differentiation of $S_{12}$, see Eq.~(\ref{deriv}) with $\tau_\mathrm{c}\simeq0.5$. 
   \textbf{a,b,c}: Data and theory at $T\simeq $\SI{11}{mK} for a source voltage $V_\mathrm{s}=\,$\SI{23}{\uV}, \SI{35}{\uV}, and \SI{70}{\uV}, respectively.
   }
   \label{FigSpectroscopy}
\end{figure*}

The sample, shown in Fig.~\ref{FigSchemSample}c, is nanostructured from an Al(Ga)As heterostructure and measured at \SI{11}{mK} and \SI{5.2}{T}.
It consists of two source QPCs (metallic split gates colored red) located at a nominal distance $d=$ \SI{3.1}{\um} from the central `analyzer' QPC (yellow gates).
If not stated otherwise, all QPCs are set to partially (fully) reflect the outer (inner) edge channel, and the analyzer QPC is tuned to an outer edge channel transmission probability $\tau_\mathrm{c}\simeq0.5$. 
A negative voltage is also applied to the non-colored gates to reflect the edge channels at all times, as schematically depicted. 
Low-frequency current auto-correlations $\langle \delta I_1^2 \rangle$ and $\langle \delta I_2^2 \rangle$ on the left and right side, respectively, and cross-correlations $\langle \delta I_1 \delta I_2 \rangle$ across the analyzer QPC are measured simultaneously. 
In the following, the excess noises are denoted $S_{ij}\equiv\langle\delta I_i \delta I_j\rangle-\langle\delta I_i\delta I_j\rangle (V_{1,2,3,4}=0)$ with $i,j\in\{1,2\}$.

\vspace{\baselineskip}
{\large\noindent\textbf{Results}}
\small

\noindent\textbf{Electron fractionalization}

\noindent
First we need to ensure that the device is in the strong coupling regime, and to determine under which conditions the tunneling electrons are fractionalized into well-separated $e/2$ charges at the analyzer QPC.

Our straightforward approach is to inject energy into one edge channel at a source, and to probe the energy redistribution at the analyzer \cite{Altimiras_EnergyEquilibration_2010}.
Indeed, the fractionalization of the tunneling electron coincides with the emergence of charge pulses in the other, co-propagating edge channel and, consequently, with a transfer of energy.
Furthermore, the full EMP delocalization between both channels, which is specific to the strong coupling limit, also translates into an equal redistribution of energy between the channels at long distances\cite{Degiovanni_PlasmonsScattering_2010}.

In this measurement, only one source QPC is used together with the analyzer.
We inject energy into the outer channel by applying a constant dc voltage bias $V_\mathrm{s}$ across the source QPC (e.g., $V_{1}= V_\mathrm{s}/2$ and $V_{3}= -V_\mathrm{s}/2$ for the left source, Fig.~\ref{FigSchemSample}c).
The resulting electron energy distribution immediately downstream of the injection point $f_\mathrm{inj}$ takes the shape of a double step  (red inset in Fig.~\ref{FigSchemSample}a), where  $f_\mathrm{inj}(\varepsilon) = \tau_\mathrm{s}f_\mathrm{FD}(\varepsilon+eV_\mathrm{s}/2) + (1-\tau_\mathrm{s})f_\mathrm{FD}(\varepsilon-eV_\mathrm{s}/2)$, 
with 
$\tau_\mathrm{s}$ the transmission probability of outer channel electrons across the source QPC, and $f_\mathrm{FD}$ the Fermi-Dirac distribution.

The electron energy distribution spectroscopy at the analyzer is performed in the out-of-equilibrium outer edge channel by measuring the cross-correlations $S_{12}$ vs the probe voltage $V_\mathrm{p}$ that  controls the electrochemical potential of the equilibrium edge channel on the other side (e.g., $V_{2}=V_{4}=V_\mathrm{p}$).
The latter's cold Fermi distribution acts as a step filter\cite{Gabelli_step_filter,Grenier_Tomography_2011,Bisognin_Tomography_2019}, up to a $k_\mathrm{B}T\approx$ 0.1 \textmu eV rounding.
The probed out-of-equilibrium electron energy distributions $f$ displayed in Fig.~\ref{FigSpectroscopy} are computed from the measured $S_{12}$ (inset) using \cite{Blanter_Bible_2000,Gabelli_step_filter}:
\begin{equation}
f(\varepsilon=eV_\mathrm{p})\equiv\frac{1}{2} \left(1+\frac{h}{2e^2\;\tau_\mathrm{c}(1-\tau_\mathrm{c})} \frac{\partial S_{12}(V_\mathrm{p})}{e\partial V_\mathrm{p}}\right).
\label{deriv}
\end{equation}

The three panels in Fig.~\ref{FigSpectroscopy} show the evolution of $f$ with the source bias voltage $V_\mathrm{s}$, at $\tau_\mathrm{s}\simeq\tau_\mathrm{c}\simeq0.5$.
Whereas at low $V_\mathrm{s}=$ \SI{23}{\uV} (panel a) $f$ remains close to a double-step function, we observe a marked relaxation towards an intermediate shape at $V_\mathrm{s}=$ \SI{35}{\uV} (panel b) and, at high bias $V_\mathrm{s}=$ \SI{70}{\uV} (panel c), $f$ takes the shape of a broad single step closely matching the long distance prediction for the strong coupling limit (blue dashed lines). 
This last observation establishes that the present device is in the strong coupling limit.
Furthermore, since the data for \SI{70}{\uV} agrees well with the long distance prediction, this indicates that the fractionalized $e/2$ wave-packets are already well-separated at the analyzer for a bias $V_\mathrm{s}$ of \SI{70}{\uV}.
The observation at \SI{35}{\uV} of a different distribution function, still showing remnants of a double step, indicates an incomplete separation of the wave-packets up to this voltage.   
Therefore the separation into two $e/2$ wave-packets occurs for a source voltage bias within the range of \SI{35}{\uV} and \SI{70}{\uV}. In that case, the wave-packet time-width $h/eV_\mathrm{s}$\cite{Martin_WPapproachSN_1992} is smaller than the time delay between the arrival of fractionalized $e/2$ charges at the analyzer QPC $\delta t=d/v_\mathrm{n}-d/v_\mathrm{c}$.

\begin{figure*}[!htb]
\renewcommand{\figurename}{\textbf{Figure}}
\renewcommand{\thefigure}{\textbf{\arabic{figure}}}
	\centering
	\includegraphics[width=13.9cm]{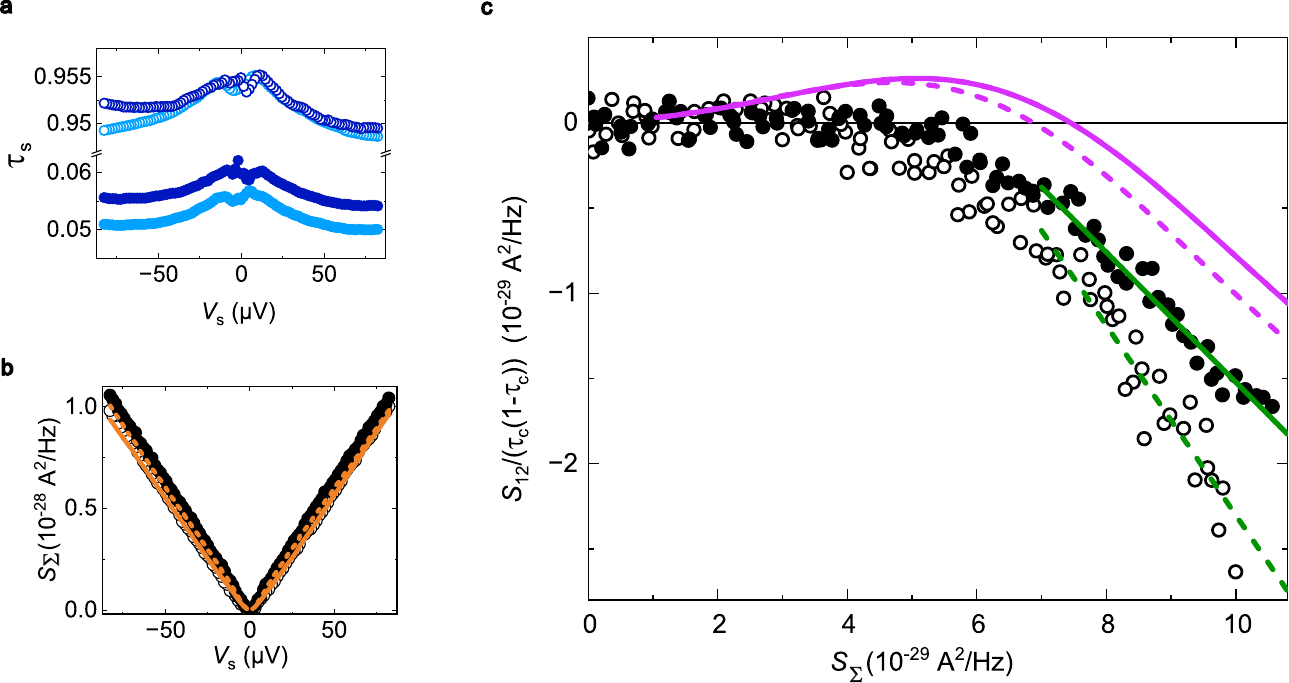}
	\caption{
	\footnotesize
	\textbf{Cross-correlation signature of fractional statistics} with symmetric dilute beams. 
    \textbf{a} Measured left/right source QPC dc transmission as a function of bias voltage, shown in light/dark blue, respectively.
    \textbf{b} Sum of sources' shot noise $S_\mathrm{\Sigma}$ vs source bias voltage $V_\mathrm{s}$.
    The orange lines display Eq.~(\ref{EqSnoint}) with $T=$ \SI{11}{mK}, the independently measured temperature.  
    \textbf{c} Measured excess shot noise $S_\mathrm{12}/(\tau_\mathrm{c}(1-\tau_\mathrm{c}))$ as a function of source shot noise $S_\mathrm{\Sigma}$ for a small source QPC transmission $\tau_\mathrm{s}=0.05/0.95$ (full/empty dots respectively).
    The purple lines display the strong inter-channel coupling prediction for $\delta t=$ \SI{64}{ps}.
    The green lines denotes the slope, i.e., the Fano factor (see Main text), yielding $P\simeq-0.38/0.56$ for $\tau_\mathrm{s}=0.05/0.95$ respectively.  
    In all panels full/open circles and solid/dashed lines denote $\tau_\mathrm{s}=0.05/0.95$ respectively.     
    }
    \label{FigCross}
\end{figure*}

Further evidence of the good theoretical description of the device is provided by the quality of the quantitative comparison between the data and the exact calculations of $f$ at finite distance (purple continuous lines).
These predictions were obtained by an extension of the theory involving a  subsequent refermionization of the bosonized Hamiltonian, which enables a full access to the cross-correlations and out-of-equilibrium electron distributions  
(see Supplementary Information).
The only fitting parameter is the time delay $\delta t$.
Here it is fixed to $\delta t=$ \SI{64}{ps}, and its associated effective velocity $d/\delta t=$\,\SI{5d4}{m.s^{-1}} is comparable to EMP velocity measurements in similar samples\cite{Bocquillon_SeparationModes_2013}.
In addition to the Supplementary Fig.~\textbf{5} showing a comparison at additional intermediate voltages, see also Supplementary Fig.~\textbf{6} for measurements with a dilute quasiparticle beam, and Supplementary Fig.~\textbf{7} for a (less-controlled) power injection in the inner edge channel.


\vspace{\baselineskip}
\noindent\textbf{Negative cross-correlation signature of anyon statistics}

\noindent
We now turn to the cross-correlation investigation of the fractional mutual braiding statistics between $e/2$ edge quasiparticles and electrons.
Figure~\ref{FigCross} displays the central measurement of cross-correlations in the configuration of two sources injecting symmetric dilute beams toward the analyzer.
The source QPCs are biased at a voltage $V_\mathrm{s}$ equally distributed on the two inputs ($V_{1,2}=-V_{3,4}=V_\mathrm{s}/2$, Fig.~\ref{FigSchemSample}c), and set to $V_\mathrm{s} = $ \SI{70}{\uV}, previously established to correspond to the full fractionalization of the quasiparticles entering the analyzer. We set both source QPCs  either to a transmission $\tau_\mathrm{s}\approx0.05$ corresponding to a dilute beam of electrons, or to  $\tau_\mathrm{s}\approx0.95$ for a dilute beam of holes (Fig.~\ref{FigCross}a).

The relevant parameter to investigate the cross-correlation signature of anyonic statistics is the generalized Fano factor \cite{Rosenow_Collider_2016} 
\begin{equation}
    P\equiv\frac{S_{\mathrm{12}}}{\tau_\mathrm{c} (1-\tau_\mathrm{c}) S_\mathrm{\Sigma}},
    \label{DefP}
\end{equation}
\noindent
where $\tau_\mathrm{c}$ is the  
analyzer transmission, and $S_\mathrm{\Sigma}$ is the sum of the current noises emitted from the two source QPCs. 
$P$ carries information on the braiding statistics, with a high bias voltage limit that depends on the braiding phase \cite{Rosenow_Collider_2016, Thamm_2023,Iyer_2023}.
If the (mutual) braiding statistics of free quasiparticles is trivial, such as for fermions or bosons, then $P$ is zero, whereas it is non-zero otherwise.
The mere observation of a non-zero $P$ therefore provides a qualitative signature of an unconventional braiding statistics.
However, we stress the importance of unambiguously establishing the underlying theoretical description of the system. For instance, negative cross-correlations could also be obtained with fermions, by phenomenologically introducing an ad hoc redistribution of energy (see Supplementary Information).  
Here we have established the suitability of the fractionalized charge picture through electron energy distribution spectroscopy, by comparing the observed bias voltage evolution of the energy distribution with the quantitative predictions of this model (see Fig.~\ref{FigSpectroscopy} and Supplementary Fig.~\textbf{5}).
In practice, $P$ is extracted from the slope of $S_\mathrm{12}/(\tau_\mathrm{c}(1-\tau_\mathrm{c}))$ vs $S_\mathrm{\Sigma}$, as shown in  Fig.~\ref{FigCross}c (green lines). 
Note that the measurements of $S_\mathrm{\Sigma}$ and $S_\mathrm{12}$ are performed simultaneously, by exploiting the current conservation relation $S_\mathrm{\Sigma}=S_{11}+S_{22}+2S_{12}$.

\begin{figure*}[!htb]
\renewcommand{\figurename}{\textbf{Figure}}
\renewcommand{\thefigure}{\textbf{\arabic{figure}}}
	\centering
	\includegraphics[width=17.9cm]{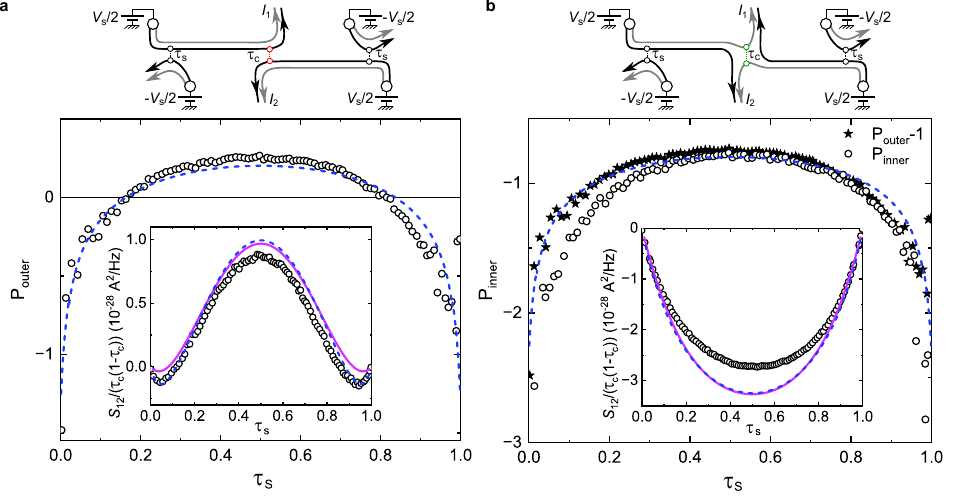}
	\caption{
	\footnotesize
	\textbf{Cross-correlations vs dilution} of symmetric beams.
 Main panels and insets show, respectively, the generalized Fano factor $P$ and the renormalized cross-correlations $S_\mathrm{12}(V_\mathrm{s}=$\SI{70}{\uV}$)/(\tau_\mathrm{c}(1-\tau_\mathrm{c}))$ vs the outer edge channel transmission $\tau_\mathrm{s}$ of the symmetric source QPCs.
 Symbols are data points.
 Blue lines are high bias/long $\delta t$ predictions.
 Purple lines are $S_\mathrm{12}(V_\mathrm{s}=$\SI{70}{\uV}$)/(\tau_\mathrm{c}(1-\tau_\mathrm{c}))$ predictions at $\delta t=$ \SI{64}{ps}.
  	\textbf{a}, The cross-correlation signal and corresponding $P_{\mathrm{outer}}$ (open circles) are measured by partially transmitting at the central QPC ($\tau_\mathrm{c}\approx0.5$) the same outer edge channel (black) where electrons are tunneling at the sources (see schematics).
   This is the standard `collider' configuration.
   \textbf{b}, The cross-correlation signal and corresponding $P_{\mathrm{inner}}$ are obtained by setting the central QPC to partially transmit ($\tau_\mathrm{c}\approx0.5$) the inner edge channel (grey), whereas electrons are tunneling into the outer edge channel at the sources (see schematic).
   In this particular configuration, the source shot noise does not directly contribute to the cross-correlation signal.
   Filled symbols in the main panel display $P_\mathrm{outer}-1$, with $P_\mathrm{outer}$ the data in (\textbf{a}) and $-1$ corresponding to the subtraction of the source shot noise.
 	}	\label{FigBlob}
\end{figure*}

We first check that $S_\mathrm{\Sigma}$ reflects the charge $e$ of injected electrons (Fig.~\ref{FigCross}b).
This is attested by the good quantitative agreement, without any fit parameter, between the data (symbols) and the shot noise prediction for electrons (orange lines) given by\cite{Blanter_Bible_2000}: 
\begin{equation}
   S_\mathrm{\Sigma} = 2 \frac{e^2}{h}\! \!\sum_{i=L,R}\!\! \tau_\mathrm{i}(1-\tau_\mathrm{i})e V_\mathrm{s}\!\left[ \coth\!\left (\frac{eV_\mathrm{s}}{2k_{\mathrm{B}}T}\right)-\frac{2k_{\mathrm{B}}T}{eV_\mathrm{s}}\right]\!\!,
    \label{EqSnoint}
\end{equation}
\noindent
with $T=$ \SI{11}{mK} and $\tau_\mathrm{L(R)}$ the measured dc transmission of the left (right) source shown in Fig.~\ref{FigCross}a. 

We then focus on the cross-correlation investigation of anyonic behavior.
As shown in Fig.~\ref{FigCross}c, $S_{12}\approx0$ at low bias, up to 
$S_\Sigma\approx$ \SI{6d-29}{A^2.Hz^{-1}}
corresponding to $|V_\mathrm{s}|\approx$ \SI{45}{\uV}.
This $P\approx0$ signals a trivial mutual statistics, which is expected in the low bias regime where the injected electron is not fractionalized at the analyzer.
Then, at $|V_\mathrm{s}|\gtrsim$ \SI{45}{\uV} where the fractionalization takes place according to $f(\varepsilon)$ spectroscopy, $S_{12}$ turns negative with a slope of $P\simeq-0.38/0.56$ (green solid/dashed line) for $\tau_\mathrm{s} = 0.05/0.95$ respectively.
The clear negative signal with a fixed slope constitutes a strong qualitative marker of non-trivial mutual braiding statistics, as further discussed below.

The relationship between negative cross-correlations and anyonic mutual statistics in the dilute limit of small $\tau_\mathrm{s}$ (or, symmetrically, small $1-\tau_s$) is most clearly established in a perturbative analysis along the lines of Morel \textit{et al}\cite{Morel_IntegerCollider_2022}.
For $\tau_\mathrm{s}\ll1$ and at long distances from the source, we find (see Supplementary Information):
\begin{equation}
    P\simeq\frac{\sin^2\theta}{\theta^2}\ln{\tau_\mathrm{s}},
    \label{eq:PMorel}
\end{equation}
with $2\theta$ the mutual braiding (double exchange) phase.
For quasiparticles of charges $q$ and $q'$ along an integer quantum Hall channel, theory predicts $\theta=\pi qq'/e^2$ (see, e.g., Ref.~\citenum{Lee_MutualStats_2020}).
In the present case of a braiding between incident fractional charges $e/2$ and spontaneously generated electron-hole pairs, we thus have $\theta=\pi/2$ and $P\simeq\frac{4}{\pi^2}\ln{\tau_\mathrm{s}}$ (see also Ref.~\citenum{Idrisov_ColliderTwo_2022} for the same prediction, but without the explicit connection to the fractional mutual statistics).
Note that in the present integer quantum Hall implementation, the relationship between $P$ and $\theta$ is not complicated by additional parameters, such as the fractional quasiparticles' scaling dimension and topological spin that both come into play in the fractional quantum Hall regime\cite{Rosenow_Collider_2016,Lee_NonAbelianCollider_2022, Iyer_2023, Thamm_2023}.
However, achieving $P\propto\ln\tau_\mathrm{s}$ requires large $|\ln\tau_\mathrm{s}|$ and thus exponentially small $\tau_\mathrm{s}$, which complicates a quantitative comparison of experimental data with Eq.~(\ref{eq:PMorel}).
Accordingly, injecting $\tau_\mathrm{s}=0.05$ ($\ln\tau_\mathrm{s}\simeq-3$) into Eq.~(\ref{eq:PMorel}) gives a slope $P\simeq-1.2$, substantially more negative than the observations.

A better data-theory agreement can be obtained with a non-perturbative treatment of the sources.
Indeed, the predicted slope in the high bias/long distance limit at $\tau_\mathrm{s}=0.05$ is $P\simeq-0.35$ (see Eq.~(52) in Supplementary Information), close to one of the observed values $P\simeq-0.38$. 
Although we expect the same cross-correlations and Fano factor for $\tau_\mathrm{s} = 0.05$ and 0.95, the measured value at 0.95 of $P \simeq-0.56$ (green dashed line) is somewhat higher, and also deviates more from the theory prediction (purple dashed line). We discuss possible reasons, such as increased inter-channel tunneling, in the Supplementary Information. The full finite bias/finite distance predictions (purple lines in Fig.~\ref{FigCross}c) also reproduce the overall shape of the measurements, although with a noticeable horizontal shift (see Supplementary Information for a discussion of possible theoretical limitations).
Further evidence of the underlying anyonic mechanism is provided from the effect of the dilution of the quasiparticle beam.

\vspace{\baselineskip}
{\noindent\textbf{Cross-correlations vs beam dilution}}

\noindent
Here we explore the effect of dilution by sweeping the transmission across the two symmetric sources over the full range $\tau_\mathrm{s}\in[0,1]$, and we also extend our investigation to the inner edge channel where the electron fractionalization results in two pulses carrying opposite charges $\pm e/2$ (see Fig.~\ref{FigSchemSample}a). 

Let us first consider the previous/standard configuration, with source QPCs and analyzer QPC set to partially transmit the same, outer, channel.
Away from the dilute limits, $P$ and $S_\mathrm{12}$ show a change of sign (see Fig.~\ref{FigBlob}a).
This results from an increasing importance of the positive contribution from the noise generated at the sources \cite{Ota_NegativeCrossNu1_2017} with respect to the noise generated at the analyzer involving the emergence of mechanisms other than time braiding (such as collisions) \cite{Morel_IntegerCollider_2022,Iyer_2023}.
In addition we see that the data are symmetric around $\tau_\mathrm{s}=0.5$, due to the unchanging electron nature of tunneling particles into the IQH edges (with small deviations possibly from inter-channel tunneling, as previously mentioned).
This is in contrast with the fractional quantum Hall regime where the nature of the tunneling quasiparticles changes \cite{Moty_QPC} between $\tau_\mathrm{s}\ll1$ and $1-\tau_\mathrm{s}\ll1$.  
The agreement with theory observed for $P(\tau_\mathrm{s})$, and more specifically for the $\tau_\mathrm{s}$ dependence in the dilute limits $\tau_\mathrm{s}\ll1$ and $1-\tau_\mathrm{s}\ll1$, further establishes the experimental cross-correlation signature of fractional mutual statistics.
Note that the less precise agreement with the full $S_{12}$ signal is reminiscent of the horizontal shift of the negative slope in Fig.~\ref{FigCross}.

We then consider in Fig.~\ref{FigBlob}b the alternative configuration, where the analyzer is set to $\tau_\mathrm{c}\simeq0.5$ for the inner edge channel (the outer edge channel, where electrons are injected at the sources, being fully transmitted, see schematics).
In that case, the cross-correlation signal is always negative due to charge conservation. 
The positive contribution $\tau_\mathrm{c}(1-\tau_\mathrm{c})S_\Sigma$ from the partition of the current noise generated at the sources is absent since the electron tunneling at the sources does not take place in the probed inner edge channel.
In the strong coupling limit where fractional charges of identical amplitude $e/2$ propagate on both inner and outer channel, the same unconventional braiding is expected to have the same cross-correlation consequences.
The absence of source noise then simply results in an offset: $P_\mathrm{inner}=P_\mathrm{outer}-1$ ($S_{12}^\mathrm{inner}=S_{12}^\mathrm{outer}-\tau_\mathrm{c}(1-\tau_\mathrm{c})S_\Sigma$), where the label inner/outer indicates the channel probed at the analyzer.
For a direct comparison, $P_\mathrm{outer}-1$ is also shown (filled stars).
The agreement between the two data sets provides an additional, direct confirmation that the device is in the strong coupling regime.
It also experimentally establishes the robust contribution from the source noise to $S_{12}$, allowing to distinguish it from the effect of time-braiding.

\vspace{\baselineskip}
{\large\noindent\textbf{Discussion}}
\small

\noindent
Edge excitations are not characterized by topologically protected quantum numbers, in contrast to the quantum Hall bulk quasiparticles.
Along the integer and fractional quantum Hall edges, the charge and the quantum statistics of such excitations can be varied continuously.
In the present work, we exploit this property to form dilute beams of fractional charges which behave as anyons.
This is achieved by using QPCs as electron sources in combination
with the intrinsic Coulomb interaction between co-propagating integer quantum Hall channels.
We establish their fractional quantum statistics by the emergence of negative current cross-correlations between the two outputs of a downstream analyzer QPC, similarly to previous observations in the fractional quantum Hall regime.
By contrast, when applying sufficiently low source bias voltages such that the tunneling electrons do not fractionalize, the absence of a cross-correlation signal coincides with their fermionic character.

We believe that the demonstrated integer quantum Hall platform opens a promising practical path to explore the emerging field of anyon quantum optics \cite{Carrega_AnyonInterfReview_2021}.
Advanced and time-resolved quantum manipulations of anyons are made possible by the large quantum coherence along the integer quantum Hall edge and the robustness of the incompressible bulk.
By tailoring single-quasiparticle wave-packets, for example with driven ohmic contacts, a vast range of fractional anyons of arbitrary exchange phase becomes available along the integer quantum Hall edges, well beyond the odd fractions of $\pi$ of Laughlin quasiparticles encountered in the fractional quantum Hall regime.


\vspace{\baselineskip}
{\large\noindent\textbf{Methods}}
\small

{\noindent\textbf{Sample fabrication}}

\noindent
The device, shown in Fig.~\ref{FigSchemSample}c of the Main text, is patterned in an AlGaAs/GaAs heterostructure forming a two-dimensional electron gas (2DEG) buried \SI{95}{\nm} below the surface. 
The 2DEG has a mobility of \SI{2.5e6}{\square\cm\per\volt\per\second} and a density of \SI{2.5e11}{\per\square\cm}.
It was nanofabricated following five standard e-beam lithography steps:
\begin{enumerate}
    \item Ti-Au alignment marks are first deposited through a PMMA mask.
    \item The mesa is defined by using a ma-N 2403 protection mask and by wet-etching the unprotected parts in a solution of H$_3$PO$_4$/H$_2$O$_2$/H$_2$O over a depth of $\sim$\SI{100}{\nm}.
    \item The ohmic contacts allowing an electrical connection with the buried 2DEG are realized by the successive depositions of Ni (\SI{10}{\nm}) - Au (\SI{10}{\nm}) - Ge (\SI{90}{\nm}) - Ni (\SI{20}{\nm}) - Au (\SI{170}{\nm}) - Ni (\SI{10}{\nm}) through a PMMA mask, followed by a 440°C annealing for \SI{50}{s}.
    \item The split gates controlling the QPCs consist in \SI{40}{\nm} of aluminium deposited through a PMMA mask.
    \item Finally, we deposit thick Cr-Au bonding ports and large-scale interconnects through a PMMA mask. 
\end{enumerate}
The nominal tip-to-tip distance of the Al split gates used to define the QPCs is \SI{150}{\nm}.


\vspace{\baselineskip}
{\noindent\textbf{Measurement setup}}\\
\noindent
The sample is installed in a cryofree dilution refrigerator with important filtering and thermalization of the electrical lines, and immersed in a perpendicular magnetic field $B=$ \SI{5.2}{T}, which corresponds to the middle of the $\nu=2$ plateau. 
Cold $RC$ filters are mounted near the device: \SI{200}{\kilo\ohm} - \SI{100}{\nano\farad} on the lines controlling the split gates, \SI{10}{\kilo\ohm} - \SI{100}{\nano\farad} on the injection lines and \SI{10}{\kilo\ohm} - \SI{1}{\nano\farad} on the low frequency measurement lines.

Lock-in measurements are made at frequencies below \SI{25}{\hertz}, using an ac modulation of rms amplitude below $k_\mathrm{B}T/e$. 
We calculate the dc currents and QPC transmissions by integrating the corresponding lock-in signal vs the source bias voltage (see the following and Ref.~\citenum{Glidic_Andreev_2023} for details).

The auto- and cross-correlations of the currents $I_\mathrm{1}$ and $I_\mathrm{2}$ (Fig.~\ref{FigSchemSample}c) are measured with home-made cryogenic amplifiers \cite{Liang_HEMTs_2012} around \SI{0.86}{MHz}, the resonant frequency of the two identical tank circuits along the two amplification chains. 
The measurements are performed by integrating the signal over the bandwidth of $[0.78,0.92]$ \SI{}{MHz}. 
The measurement setup is detailed in the supplemental material of Ref.~\citenum{Jezouin_QLimHeatFlow_2013b}.

\vspace{\baselineskip}
{\noindent\textbf{Thermometry}}\\
\noindent
The electron temperature in the sample is measured using the robust linear dependence of the thermal noise $S(T)\propto T$.
At $T>$ \SI{40}{\milli\kelvin}, we use the (equilibrium) thermal noise plotted versus the temperature readout by the calibrated RuO$_2$ thermometer. 
The linearity is a confirmation of the electron thermalization and of the thermometer calibration.
The quantitative value of the slope provides us with the gain of the full noise amplification chain, as detailed in the next section. 
To determine the temperature in the $T<$ \SI{40}{\milli\kelvin} range, we measure the thermal noise and determine the corresponding temperature by linearly extrapolating from the $S(T>$ \SI{40}{\milli\kelvin}$)$ data. 
The values of $T$ obtained using the two amplification chains are found to be consistent. 
We also check that $T$ corresponds to the temperature obtained from standard shot noise measurements performed individually on each QPC ahead of and during each measurement. A \SI{1}{\milli\kelvin} higher shot noise temperature is specifically associated with the top-right ohmic contact feeding the right source, and attributed to noise from the corresponding connecting line.

\vspace{\baselineskip}
{\noindent\textbf{Calibration of the noise amplification chain}}\\
\noindent
For each noise amplification chain $i\in\{1,2\}$, the gain factor $G^\mathrm{eff}_\mathrm{i}$ between current noise spectral density and raw measurements needs to be calibrated.
From the slopes $s_1$ and $s_2$ of $S(T>$ \SI{40}{\milli\kelvin}$)$ measured, respectively, for the amplification chain 1 and 2 (see Thermometry), and the robust fluctuation-dissipation relation $S(T)=4k_\mathrm{B}T\,\mathrm{Re}[Z]$ with $Z$ the frequency dependent impedance of the tank circuit in parallel with the sample, we get:
\begin{equation}
    G_\mathrm{1(2)}^\mathrm{eff}=\frac{s_\mathrm{1(2)}}{4k_\mathrm{B}(1/R_\mathrm{tk}^\mathrm{1(2)}+1/R_\mathrm{H})},
    \label{EqGeff}
\end{equation}
\noindent with $R_\mathrm{H}=h/2e^2$ the Hall resistance of the sample, and $R_\mathrm{tk}^\mathrm{1(2)}=$ \SI{150}{\kilo\ohm} (\SI{153}{\kilo\ohm}) the separately obtained effective parallel resistance due to the dissipation in the tank circuit connected to the same port.
With $G_\mathrm{1}^\mathrm{eff}$ and $G_\mathrm{2}^\mathrm{eff}$ given by the above relation, the gain for the cross-correlation signal reads $G_\mathrm{12}^\mathrm{eff} = \sqrt{G_\mathrm{1}^\mathrm{eff}G_\mathrm{2}^\mathrm{eff}}$ thanks to the good match of the two resonators.
For more details see Ref.~\citenum{Glidic_Andreev_2023}.

\vspace{\baselineskip}
{\noindent\textbf{Differential (ac) and integral (dc) transmission}}
\noindent

\noindent \underline{Source transmissions:}

\noindent The transmissions of the left and right source QPC $\tau_\mathrm{L,R}$ are defined as the ratio between the dc current transmitted across the QPC 

\begin{equation}
    I_\mathrm{L,R} = \int_0^{V_{3,4}} \frac{\partial (I_1+I_2)}{\partial V_{3,4}}dV_{3,4}
\end{equation}

\noindent and the dc current $e^2V_{3,4}/h$ incident on one side of the QPC for the considered outer edge channel (i.e., half of the total injected current), yielding: 

\begin{equation}
    \tau_\mathrm{L,R} =  \frac{2h}{e^2}\frac{I_\mathrm{L,R}}{ V_{3,4}}.
\end{equation}

In the specific case where the sources are set to partial transmission of the inner channel (case of  Fig.~\textbf{7} in the Supplementary Information), source transmissions of the inner channel can similarly be defined as

\begin{equation}
    \tau_\mathrm{L,R}^\mathrm{\,\,\,\,\,\,\,\,in} =  \frac{2h}{e^2 V_{1,2}} \int_{0}^{V_{1,2}} \left( \frac{e^2}{h}-\frac{\partial(I_1+I_2)}{\partial V_{1,2}}\right)dV_{1,2}.
\end{equation}

\noindent \underline{Analyzer transmission:}
Whereas above we use dc transmissions for the sources, in all measurements the relevant transmission of the analyzer is the ac transmission, obtained as  
\begin{equation}
\tau_\mathrm{c} =\left. \left( \frac{\partial I_2}{\partial V_3} \right) \middle/ \left(\frac{\partial (I_1 + I_2)}{\partial V_3}\right)\right.. 
\end{equation}
In particular, there is no dc bias of the analyzer in the symmetric source configuration probing the quantum statistics.

\vspace{\baselineskip}
{\large\noindent\textbf{Data and code availability}}
\small

\noindent
Plotted data, raw data, data analysis code and numerical code used to calculate the full theoretical predictions have been deposited in Zenodo under the accession code:  https://doi.org/10.5281/zenodo.10492057. 
\normalsize

\vspace{\baselineskip}
{\large
 

\vspace{\baselineskip}
\small
{\large\noindent\textbf{Acknowledgments}}\\
We would like to thank P.\ Degiovanni and J. T.\ Chalker for discussions. 
This work was supported by the European Research Council (ERC-2020-SyG-951451), the French National Research Agency (ANR-16-CE30-0010-01 and ANR-18-CE47-0014-01), the French RENATECH network and DIM NANO-K.
D.K.\ acknowledges support from Labex MME-DII grant ANR11-LBX-0023, and funding under The Paris Seine Initiative EMERGENCE programme 2019.

\vspace{\baselineskip}
{\large\noindent\textbf{Author Contributions}}\\
P.G., C.P.\ and I.P.\ performed the experiment and analyzed the data with inputs from A.Aa., A.An., D.K.\ and F.P.; 
D.K.\ and C.M.\ developed the theory; 
D.K.\ wrote the code for numerical predictions; 
A.C.\ and U.G.\ grew the 2DEG; 
P.G\ and A.Aa.\ fabricated the sample with inputs from A.An.; 
Y.J.\ fabricated the HEMTs used in the cryogenic amplifiers for noise measurements; 
I.P.\ and P.G.\ wrote the manuscript with contributions from all authors; 
A.An.\ and F.P.\ led the project.

\vspace{\baselineskip}
{\large\noindent\textbf{Competing interests}}\\
The authors declare no competing interests.

\vspace{\baselineskip}
{\large\noindent\textbf{Additional Information}}\\
\textbf{Correspondence} and requests for materials should be addressed to I.P.\ and F.P.


\clearpage

\onecolumngrid
\begin{center}
\vspace{4mm}
\large{\textbf{ Supplementary Information \\
for \\
Signature of anyonic statistics in the integer quantum Hall regime }}
\end{center}

\setcounter{figure}{0}

\maketitle

\vspace{5mm}

\textit{Note that the Supplementary Information refers to its own set of references, separate from those in the main manuscript.}

\section{Non-perturbative theory}

\subsection{Introduction}
In this Section we outline the theoretical description of the electron collider using a model which can be solved via refermionization techniques. 
We also present the results for the asymptotics of the noise in the small tunneling limit.

The notation in the `Non-perturbative theory' Section is different from the rest of the paper in order to keep the correspondence with the previous work Ref.~\citenum{kovrizhin2012} and with an upcoming theoretical publication providing further details (D.~Kovrizhin, in preparation). 

The notation conversion is the following : 
\begin{table}[h!]
  \begin{center}
\renewcommand{\tablename}{\textbf{Supplementary Table}}
    \label{tab:table1}
    \begin{tabular}{c|c}
      \textbf{    This section    } & \textbf{  Elsewhere    } \\
      \hline
      $T_1$ & $\tau_\mathrm{s}$ \\
      $T_2$ & $\tau_\mathrm{s}$ \\
      $T_S$ & $\tau_\mathrm{c}$ \\
      bias $V$ & bias $V_\mathrm{s}$ \\
      spin up ($\uparrow$) channel & inner channel \\
      spin down ($\downarrow$) channel & outer channel \\
      1' channels (Fig. \ref{fig:collider})  & region biased by $V_3$ (Fig. \textbf{1}c Main text) \\
      2' channels (Fig. \ref{fig:collider}) & region biased by $V_1$ (Fig. \textbf{1}c Main text) \\
      1 channels (Fig. \ref{fig:collider}) & region biased by $V_2$ (Fig. \textbf{1}c Main text) \\
      2 channels (Fig. \ref{fig:collider}) & region biased by $V_4$ (Fig. \textbf{1}c Main text) \\
    \end{tabular}
    \caption{Notation correspondence between this Section and the rest of the paper.}  
  \end{center}
\end{table}

The schematics of the model is shown in Supplementary Fig.~\ref{fig:collider} for the noise measurement between channels $1_\downarrow$ and $2'_\downarrow$. In the Main text this is outlined as the default configuration, i.e., injection and measurement in the outer channel. 
Other configurations can be calculated in a similar manner, and the theory will be detailed in the above mentioned upcoming article. 

Let us consider a system with four edge states $1,1',2,2'$, each of which is carrying two co-propagating edge channels ($\uparrow,\downarrow$), where the arrows denote the spin. The biasing scheme of biasing source quantum point contacts with 0 and $eV$ in Supplementary Fig.~\ref{fig:collider} is equivalent to biasing with $-eV/2$ and $+eV/2$ (used in the experiment) due to invariance of the observables under a global potential shift.

\begin{figure*}[!htb]
\renewcommand{\figurename}{\textbf{Supplementary Figure}}
\renewcommand{\thefigure}{\textbf{\arabic{figure}}}
	\centering
	\includegraphics[width=12cm]{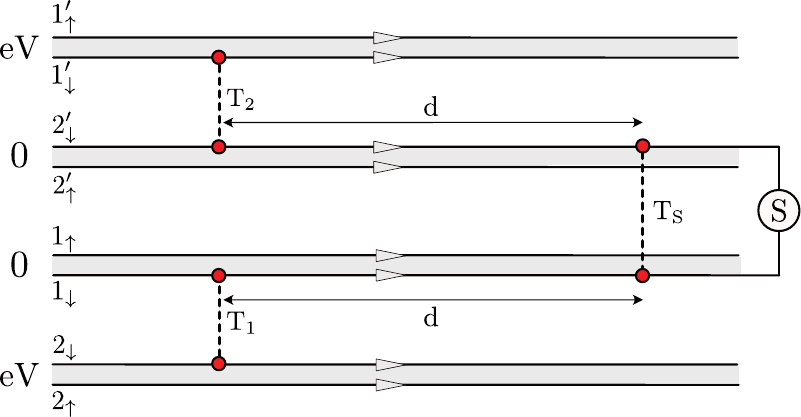}
	\caption{
	\footnotesize
	Schematic of the model of the electron collider at $\nu=2$ with four chiral quantum Hall edges (1, $1'$, 2, $2'$), each carrying two edge channels ($\uparrow$, $\downarrow$). 
 The channels $1_\downarrow$ and $2_\downarrow$ are coupled via the source QPC$_1$ positioned at $x=0$ and characterized by the tunneling probability $T_1$.
 Similarly the edge states $1'_\downarrow$ and $2'_\downarrow$ are coupled by QPC$_2$ with tunneling probability $T_2$.
 At $x=d$ downstream from the sources, the channels $2'_\downarrow$ and $1_\downarrow$ are coupled by the analyzer QPC$_S$ of tunneling probability $T_S$.
 The considered interaction is short-range, acting between the channels which propagate along the same quantum Hall edge, separated by a narrow strip shown in grey. 
 The two top channels ($1'_{\uparrow,\downarrow}$) and the two bottom channels ($2_{\uparrow,\downarrow}$) are biased with the chemical potentials $eV$, whereas the four channels in the middle ($1_{\uparrow,\downarrow}$, $2'_{\uparrow,\downarrow}$) are grounded.
 The current correlations are measured after QPC$_S$. 
Solvable variants of the model include using the analyzer QPC$_S$ to couple the other channels ($2'_\uparrow$ with $1_\uparrow$) for the data-theory comparison in, e.g.,\ Fig.~\textbf{4}b of the Main text, or applying the same voltage $V_\mathrm{p}$ to the four bottom channels ($1_{\uparrow,\downarrow}$, $2_{\uparrow,\downarrow}$) for the data-theory comparison on electron distribution spectroscopy (see corresponding schematic in Supplementary Fig.~\ref{fig:distri}).
 }	\label{fig:collider}
\end{figure*}

We model the non-interacting edge channels by the free-fermion Hamiltonian with linear dispersion (assuming the same Fermi-velocity $v_F$ in every channel). 
This is a standard description of edge states at integer filling factors. 
In addition, we assume that electrons on each edge interact via short-range interactions with strength $g$ which has the dimension of a velocity. 
This model of interactions provides a good description of previous experiments at filling factor $\nu=2$, where interactions are usually sufficiently strong to overcome any asymmetry between channels \cite{Kovrizhin2010,kovrizhin2011,kovrizhin2012}. 
We note that in the experimental setup the long-range Coulomb interactions are expected to be screened by the metallic gates over a typical length scale of a few hundred nanometers of the order of the distance between 2DEG and nearest gate. 

The Hamiltonian of the system shown in Supplementary Fig.~\ref{fig:collider} reads:
\begin{multline}\label{ham_ferm}
\hat{H}=-i\hbar v_F\sum_{\eta=1,1',2,2',s=\uparrow,\downarrow}\int_{-\infty}^{\infty}dx\ \hat{\Psi}^{\dagger}_{\eta s}(x)\partial_x\hat\Psi_{\eta s}(x)+2\pi\hbar g\sum_{\eta=1,1',2,2'}\int_{-\infty}^{\infty}\hat{\rho}_{\eta\uparrow}(x)\hat{\rho}_{\eta\downarrow}(x)dx\\+
\left(v_1 \hat{\Psi}^{\dagger}_{1\downarrow}(0)\hat\Psi_{2\downarrow}(0) + v_2 \hat{\Psi}^{\dagger}_{1'\downarrow}(0)\hat\Psi_{2'\downarrow}(0)+
v_S \hat{\Psi}^{\dagger}_{2'\downarrow}(d)\hat\Psi_{1\downarrow}(d) +\mathrm{h.c.}\right).
\end{multline}
Here $d$ is the distance between the source QPC$_{1,2}$ and the analyzer QPC$_S$.
$\hat{\Psi}_{\eta s}(x)$ are fermion annihilation operators and $\hat{\rho}_{\eta s}(x)$ are fermion density operators on the corresponding edge channels ($\eta s$).
$v_{1,2,S}$ are the tunneling amplitudes connected to the transmission probabilities $T_{1,2,S}$ across QPC$_{1,2,S}$ ($T_{1,2,S}=\sin^2(|v_{1,2,S}|/\hbar v_F)$ in the non-interacting case).
See Ref.~\citenum{kovrizhin2013} for details of the refermionization approach with the same notations as in this Section. 

The zero-frequency cross-correlation noise $S_{1\downarrow 2'\downarrow}$ between the channels $1_\downarrow$ and $2'_\downarrow$ outgoing from the analyzer QPC$_S$, at the position $x\geqslant d$ (see Supplementary Fig.~\ref{fig:collider}), is obtained from:
\begin{equation}
S_{1\downarrow 2'\downarrow}(V)=2\int_{-\infty}^{\infty}dt\ \langle\delta\hat I_{1\downarrow}(x,t)\delta\hat I_{2'\downarrow}(x,0)\rangle,
\label{eq_s12_def}
\end{equation}
with $\delta\hat I_{\eta s}(x,t)$ written in terms of the current operators $I^H_{\eta s}(x,t)$ in the Heisenberg representation as:
\begin{equation}
\delta\hat I_{\eta s}(x,t)=\hat I^H_{\eta s}(x,t)-\langle\hat I^H_{\eta s}(x,t)\rangle.
\end{equation}
The current operators in the interaction representation, where the tunneling plays the role of the `interaction', are given by the following expression
\begin{equation}
\hat{I}_{\eta s}(x,t)=-e v_F \hat\rho_{\eta s}(x,t).
\end{equation}

\vspace{\baselineskip}
The model described by the Hamiltonian given in Eq.~(\ref{ham_ferm}) can be solved for arbitrary interaction strengths using the refermionization approach\cite{kovrizhin2013}, as sketched in the next subsections (further details in D.~Kovrizhin, in preparation).
As a check, note that when applying this refermionization approach to the non-interacting case $g=0$ at $T=0$, we recover the same expression derived in the free electron scattering model:
\begin{equation}
S^{(0)}_{1\downarrow 2'\downarrow}(V)=-2 \left(\frac{e^2}{2\pi\hbar}\right) T_S R_S (T_1-T_2)^2 |eV|,
\label{s12_nonint}
\end{equation}
where $R_{1,2,S}\equiv 1-T_{1,2,S}$ are the reflection probabilities.
Note also that the noise in the non-interacting case is independent of the distance $d$, and is always zero in the case of equal tunneling probabilities on the source QPC$_{1,2}$ ($T_1=T_2$).

\subsection{Bosonization}
We would like to bosonize the Hamiltonian in Eq.~(\ref{ham_ferm}).
After introducing bosonic operators $\hat\phi_{\eta s}(x)$, Klein factors $\hat F_{\eta s}$, and number operators $\hat N_{\eta s}$ in the Schr\"{o}dinger representation, as well as a short-distance cutoff $a$, we can write the fermion operators in the bosonic form:
\begin{equation}
\hat{\Psi}_{\eta s}(x)=\frac{1}{\sqrt{2 \pi a}}\hat F_{\eta s} e^{i \frac{2\pi}{L}\hat N_{\eta s}x}e^{-i\hat{\phi}_{\eta s}(x)}.
\end{equation}
The operators obey the following commutation relations:
\begin{equation}
\lbrack \hat{\phi}_{\eta s}(x),\partial _{y}\hat{\phi}_{\eta ^{\prime} s ^{\prime}}(y)]=-2\pi
i\delta _{\eta \eta ^{\prime }}\delta _{ss^{\prime }}\delta (x-y),\ \ \{\hat{F}_{\eta s}^{\dagger },\hat{F}_{\eta ^{\prime }s^{\prime }}\}=2\delta
_{\eta \eta ^{\prime }}\delta _{ss^{\prime }},\text{ \ }[\hat{N}_{\eta s},%
\hat{F}_{\eta ^{\prime }s^{\prime }}]=-\delta _{\eta \eta ^{\prime }}\delta_{ss^{\prime}}\hat{F}_{\eta s},
\end{equation}
and the density operator is written in terms of the bosonic fields as
\begin{equation}  \label{density}
\hat{\rho}_{\eta s}(x)=\frac{\hat{N}_{\eta s}}{L}-\frac{1}{2\pi}\partial _{x}\hat{\phi}%
_{\eta s}(x).
\end{equation}

\noindent
In this bosonic representation, the Hamiltonian in Eq.~(\ref{ham_ferm}) can be written  as:
\begin{multline}
\hat{H}=\frac{\hbar v_{F}}{2}\sum_{\eta s}\int \frac{dx}{2\pi }(\partial
_{x}\hat{\phi}_{\eta s})^{2}+g\hbar \sum_{\eta }\int \frac{dx}{2\pi }%
\partial _{x}\hat{\phi}_{\eta \uparrow }\partial _{x}\hat{\phi}_{\eta
\downarrow }  \label{rf_h_bs}
+\frac{2\pi g\hbar }{L}\sum_{\eta }\hat{N}_{\eta \uparrow }\hat{N}_{\eta
\downarrow }+\frac{2\pi }{L}\frac{\hbar v_{F}}{2}\sum_{\eta s}\hat{N}_{\eta
s}(\hat{N}_{\eta s}+1)\\+\frac{1}{2 \pi a}\left(v_{1}\hat{F}%
_{1\downarrow }^{\dagger }\hat{F}_{2\downarrow }e^{i(\hat{\phi}_{1\downarrow
}(0)-\hat{\phi}_{2\downarrow }(0))}+v_{2}\hat{F}%
_{1'\downarrow }^{\dagger }\hat{F}_{2'\downarrow }e^{i(\hat{\phi}_{1'\downarrow
}(0)-\hat{\phi}_{2'\downarrow }(0))}+v_{S}\hat{F}%
_{2'\downarrow }^{\dagger }\hat{F}_{1\downarrow }e^{i(\hat{\phi}_{2'\downarrow
}(0)-\hat{\phi}_{1\downarrow }(0))}+\mathrm{h.c.}\right).
\end{multline}

\subsection{Refermionization}
Here we show how to refermionize the Hamiltonian in Eq.~(\ref{rf_h_bs}), 
which allows one to obtain the exact expressions for the noise in the presence of interactions.
To express the current correlators we will use refermionization, which permits us to map our model onto a system of non-interacting fermions for each QPC separately. 
This does not mean that there is no dependence of the noise on the interactions because the transformations between the new fields and the original fields will generate an interaction-dependent contribution to the noise.

We start with the refermionization of the four channels $(1_\uparrow,1_\downarrow,2'_\downarrow,2'_\uparrow)$.
For this purpose we first introduce new bosonic operators $\tilde\chi_{S+}(x),\tilde\chi_{A-}(x),\tilde\chi_{A+}(x),\tilde\chi_{S-}(x)$, which are related to the original bosonic operators $\phi_{\eta s}$ via the transformation $\tilde\chi^T=U\phi^T$ where $U$ is the following $4\times 4$ matrix:
\begin{equation}
U=\frac{1}{2}\left(\begin{array}{rrrr}
1 & 1 & 1 & 1\\
1 & -1 & 1 & -1\\
1 & 1 & -1 & -1\\
1 & -1 & -1 & 1
\end{array}\right).
\end{equation}
Note that the dispersions of the bosons corresponding to $\tilde\chi_{A,S+}$ and $\tilde\chi_{A,S-}$ are given by the velocities $v_+=v_F+g$ and $v_-=v_F-g$, respectively.  

We would like to evaluate current correlation functions at non-equal times for the channels $1_\downarrow$ and $2'_\downarrow$ at some position after the QPC$_S$.
For that we need to have expressions for the currents in these channels in terms of refermionized operators.
The currents in terms of the original fermions can be obtained from the Heisenberg equations of motion for the density operators (in the interaction representation, where the tunneling is treated as an interaction term in the Hamiltonian):
\begin{equation}
\partial_t\hat \rho_{1\downarrow}(x,t)=-\partial_x(v_{F}\hat\rho_{1\downarrow}(x,t)+g\hat\rho_{1\uparrow}(x,t)) ,\ \ \partial_t\hat \rho_{2'\downarrow}(x,t)=-\partial_x(v_{F}\hat\rho_{2'\downarrow}(x,t)+g\hat\rho_{2'\uparrow}(x,t)).
\end{equation}
The corresponding currents are expressed in terms of the original fermions in the interaction representation as
\begin{equation}
\hat I_{1\downarrow}(x,t)=-e(v_{F}\hat\rho_{1\downarrow}(x,t)+g\hat\rho_{1\uparrow}(x,t)) ,\ \ \hat I_{2'\downarrow}(x,t)=-e(v_{F}\hat\rho_{2'\downarrow}(x,t)+g\hat\rho_{2'\uparrow}(x,t)),
\end{equation}
with $\hat\rho_{\eta s}$ given by Eq.~(\ref{density}). 
In the following we will omit the number operators $\hat{N}_{\eta s}$ appearing in Eq.~(\ref{density}) in order to simplify the notations.
Since they transform in the same way as the fields under linear transformation with the matrix $U$, we will be able to restore them at the end. 

Using the inverse transformation $U^{-1}$ ($U^{-1}=U$) we can write the currents at position $d$ in terms of the transformed density operators $\tilde{\rho}_{S,A,\pm}$, which are related to the $\tilde\chi$ fields via an equation analogous to Eq.~(\ref{density}), as
\begin{eqnarray}
\hat I_{1\downarrow}(d,t)=-e \frac{1}{2}\left(v_{+}(\tilde\rho_{S+}(d,t)+\tilde\rho_{A+}(d,t))-v_{-}(\tilde\rho_{A-}(d,t)+\tilde\rho_{S-}(d,t))\right),\\
\hat{I}_{2'\downarrow}(d,t)=-e\frac{1}{2}\left(v_{+}(\tilde\rho_{S+}(d,t)-\tilde\rho_{A+}(d,t))+v_{-}(\tilde\rho_{A-}(d,t)-\tilde\rho_{S-}(d,t))\right).
\end{eqnarray}
We can rewrite these expressions in the more convenient form
\begin{equation}
\hat I_{1\downarrow}(d,t)=\frac{e}{2}(\hat I_0(d,t)+\hat I_1(d,t)),\ \  \hat I_{2'\downarrow}=\frac{e}{2}(\hat I_0(d,t)-\hat I_1(d,t)),
\end{equation}
where we have defined
\begin{equation}
\hat I_0(d,t)\equiv -v_{+}\tilde\rho_{S+}(d,t)+v_{-}\tilde\rho_{S-}(d,t), \ \ \hat I_1(d,t)\equiv -v_{+}\tilde\rho_{A+}(d,t)+v_{-}\tilde\rho_{A-}(d,t).
\end{equation}

\vspace{\baselineskip}
We can now proceed with the calculations of the noise, obtained by integrating in time the following current correlator 
\begin{equation}
\langle \hat I_{1\downarrow}(d,t_1)\hat I_{2'\downarrow}(d,t_2)\rangle=\langle \hat I_0(d,t_1)\hat I_0(d,t_2)\rangle-\langle \hat I_1(d,t_1)\hat I_1(d,t_2)\rangle-
\langle \hat I_0(d,t_1)\hat I_1(d,t_2)\rangle+\langle \hat I_1(d,t_1)\hat I_0(d,t_2)\rangle.
\end{equation}
We now proceed with the above correlator. 
Whereas the last two terms cancel out as $\hat I_0(d,t_{1,2})$ and $\hat I_1(d,t_{2,1})$ commute, it is not the case of the first two terms.
After refermionization, one can write the Hamiltonian for the four channels coupled by QPC$_S$ in terms of free fermions with the standard tunneling term 
$H^{\rm{ref}}_{T_S}=\tilde v_S \tilde\Psi^{\dagger}_{A_{+}}(d)\tilde\Psi_{A_{-}}(d)+\mathrm{h.c.}$, with $\tilde v_S$ being the renormalized tunneling strength \cite{kovrizhin2013} directly related to the transmission probability $T_S$ measured in the experiment.
Note that $H^{\rm{ref}}_{T_S}$ does not affect the fields $S_{\pm}$.
The fermion operators $\tilde\Psi_{A_{\pm}}$ are transformed by QPC$_S$ as
\begin{align}
\tilde\Psi_{A_{+}}(d^+,t)=r_S\tilde\Psi_{A_{+}}(d^-,t)-i t_S \tilde\Psi_{A_{-}}(d^-,t),\nonumber \\
\tilde\Psi_{A_{-}}(d^+,t)=-i t_S \tilde\Psi_{A_{+}}(d^-,t)+r_S\tilde\Psi_{A_{+}}(d^-,t),
\label{eq_scattering}
\end{align}
denoting the transmission and reflection amplitudes as $t_S$ and $r_S$ ($|t_S|^2\equiv T_S$, $|r_S|^2\equiv R_S\equiv 1-T_S$), and where $d^+$ and $d^-$ are the positions just after and just before QPC$_S$, respectively.

Let us start with $\langle \hat I_0(d,t_1)\hat I_0(d,t_2)\rangle$.
Because the operator $\hat I_{0}$ does not transform under the action of $H^{\rm{ref}}_{T_S}$, we can write the current $\hat I_0$ in terms of the original bosonic fields $\hat\phi$: 
\begin{multline}
\hat{I}_{0}(d,t)=-\frac{v_{+}}{2}\left(\hat\phi_{1\uparrow}(d,t)+\hat\phi_{1\downarrow}(d,t)+\hat\phi_{2'\downarrow}(d,t)+\hat\phi_{2'\uparrow}(d,t)\right)
+\frac{v_{-}}{2}\left(\hat\phi_{1\uparrow}(d,t)-\hat\phi_{1\downarrow}(d,t)-\hat\phi_{2'\downarrow}(d,t)+\hat\phi_{2'\uparrow}(d,t)\right).
\end{multline}
We note that the operators on the right hand side are given in the Heisenberg representation, which includes the interactions and the tunneling at both QPC$_1$ and QPC$_2$.  

We now need to refermionize the subsystems connected by QPC$_1$ and QPC$_2$ (e.g., we refermionize separately channels with primed indices, and channels with unprimed indices). In order to do that we introduce operators $\hat\chi(x)$ related to transformations of the bottom four channels and operators $\hat\chi'(x)$ related to top four channels
\begin{align}
(\hat\chi_{S+}(x),\hat\chi_{A-}(x),\hat\chi_{A+}(x),\hat\chi_{S-}(x))^T&=U(\hat\phi_{1\uparrow}(x),\hat\phi_{1\downarrow}(x),\hat\phi_{2\downarrow}(x),\hat\phi_{2\uparrow}(x))^T\\
(\hat\chi'_{S+}(x),\hat\chi'_{A-}(x),\hat\chi'_{A+}(x),\hat\chi'_{S-}(x))^T&=U(\hat\phi_{1'\uparrow}(x),\hat\phi_{1'\downarrow}(x),\hat\phi_{2'\downarrow}(x),\hat\phi_{2'\uparrow}(x))^T.
\end{align}
Using these transformations we can write the current operator $\hat{I}_{0}(d,t)$ as 
\begin{equation}
\hat{I}_{0}(d,t)=-\frac{v_{+}}{2}\left(\hat\rho_{S_{+}}(d,t)+\hat\rho_{A_{+}}(d,t)+\hat\rho'_{S_{+}}(d,t)-\hat\rho'_{A_{+}}(d,t)\right)
+\frac{v_{-}}{2}\left(\hat\rho_{A_{-}}(d,t)+\hat\rho_{S_{-}}(d,t)-\hat\rho'_{A_{-}}(d,t)+\hat\rho'_{S_{-}}(d,t)\right).
\end{equation}
We note that there is no coherence between primed and unprimed terms as well as between $S_+,S_-$ terms because they are not connected by refermionized QPC$_{1,2}$ (as with QPC$_S$, only $A_+, A_-$ and $A'_{+},A'_{-}$ channels are connected by QPC$_{1,2}$ correspondingly).

The operators in the current (taken after QPC$_{1,2}$) should be related to the operators before QPC$_{1,2}$.
For this purpose, we use
\begin{equation}
\hat\Psi^{\dagger}_{A_-}(0^+,\tau)\hat\Psi_{A_-}(0^+,\tau)=\left(+i t_1 \hat\Psi^{\dagger}_{A_+}(0^-,\tau)+r_1 \hat\Psi^{\dagger}_{A_-}(0^-,\tau) \right)\left(-i t_1 \hat\Psi_{A_+}(0^-,\tau)+r_1 \hat\Psi_{A_-}(0^-,\tau)\right),
\end{equation}
and the similar transformation of $A_+$ operators at $0^{+}$.
\color{black}
These transformations allow us to write the current correlators (noting that the $A_{+}$ and $A_{-}$ operators on the right hand side are incoherent because they are taken before QPC$_1$, at position $0^-$). Denoting as $\delta\tau\equiv t_1-t_2-d/v_\mathrm{eff}$, where $v_\mathrm{eff}=(1/v_- -1/v_+)^{-1}$ is an effective velocity, we obtain (for voltage-dependent terms)
\begin{multline}\label{current_1}
\langle\langle\hat{I}_{0}(d,t_1)\hat{I}_{0}(d,t_2)\rangle\rangle=R_1 T_1 \left(G_{A_+}(\delta\tau)\bar G_{A_-}(\delta\tau)+G_{A_-}(\delta\tau)\bar G_{A_+}(\delta\tau)\right)\\
+R_2 T_2\left(G_{A'_+}(\delta\tau)\bar G_{A'_-}(\delta\tau)+G_{A'_-}(\delta\tau)\bar G_{A'_+}(\delta\tau)\right),
\end{multline}
where double brackets denote normal-ordering of the current operators, 
and 
\begin{equation}
G_{S,A,\pm}(\delta \tau)=\frac{i}{2\pi}\frac{e^{\frac{i}{\hbar}\mu_{S,A,\pm}\delta \tau}}{\delta \tau-ia}
\end{equation}
are free-fermion Green functions.
In order to evaluate this correlator we need to know the chemical potentials of the channels in the refermionized representation. These chemical potentials follow the refermionization prescription (and can be obtained using matrix $U$), which gives for this voltage setup (the distribution function setting would correspond to different values, see Ref.~\citenum{kovrizhin2012})
\begin{align} 
\mu_{A_+}&=-eV,\  \mu_{A_-}=0,\  \mu_{S_+}=eV,\  \mu_{S_-}=0, \nonumber \\
\mu_{A'_+}&=eV,\ \mu_{A'_-}=0, \ \mu_{S'_+}=eV, \ \mu_{S'_-}=0.\label{muASpm}
\end{align}
This gives:
\begin{equation}\label{current_1_mu}
\langle\langle\hat{I}_{0}(d,t_1)\hat{I}_{0}(d,t_2)\rangle\rangle=-\frac{2}{(2\pi)^2}\frac{1}{\delta\tau^2} (R_1 T_1+R_2 T_2)\cos(eV\delta\tau/\hbar),
\end{equation}
where we note that the result is independent of the position $d$.

Similarly, after some algebra, we find for the correlator of $I_1$ operators the following expression (shown here again for voltage-dependent terms)
\begin{multline}\label{current_2}
\langle\langle\hat{I}_{1}(d,t_1)\hat{I}_{1}(d,t_2)\rangle\rangle=
+(R_s-T_s)^2 R_1 T_1 \left(G_{A_-}(\delta\tau)\bar G_{A_+}(\delta\tau)+G_{A_+}(\delta\tau)\bar G_{A_-}(\delta\tau)\right)\\+
(R_s-T_s)^2 R_2 T_2 \left(G_{A'_-}(\delta\tau)\bar G_{A'_+}(\delta\tau)+G_{A'_+}(\delta\tau)\bar G_{A'_-}(\delta\tau)\right)\\
+4 R_s T_s \tilde v^2\left(G_{1\downarrow}(d,t_1-t_2)\bar G_{2'\downarrow}(d,t_1-t_2)+G_{2'\downarrow}(d,t_1-t_2)\bar G_{1\downarrow}(d,t_1-t_2)\right),
\end{multline}
where we have introduced the interacting Green functions defined as 
\begin{equation}\label{GF}
    G_{1\downarrow}(d,t_1-t_2)=\langle\hat\Psi^{\dagger}_{1\downarrow}(d,t_1)\hat\Psi_{1\downarrow}(d,t_2)\rangle,\ \ \ \bar G_{2'\downarrow}(d,t_1-t_2)= \langle\hat\Psi_{2'\downarrow}(d,t_1)\hat\Psi^{\dagger}_{2'\downarrow}(d,t_2)\rangle.
\end{equation}
The first two terms in Eq.~(\ref{current_2}) involve solely the non-interacting Green functions ($G_{A_\pm}$,$G_{A'_\pm}$) and can be expressed using the values of the chemical potentials given Eq.~(\ref{muASpm}) as $2(R_S-T_S)^2 (R_1 T_1 + R_2 T_2)\cos(eV \delta\tau/\hbar)/\delta\tau^2$, in a similar form as Eq.~(\ref{current_1_mu}). 
This gives:
\begin{multline}\label{current_2_mu}
\langle\langle\hat{I}_{1}(d,t_1)\hat{I}_{1}(d,t_2)\rangle\rangle=
2(R_S-T_S)^2 (R_1 T_1 + R_2 T_2)\cos(eV \delta\tau/\hbar)/\delta\tau^2\\
+4 R_S T_S \tilde v^2\left(G_{1\downarrow}(d,t_1-t_2)\bar G_{2'\downarrow}(d,t_1-t_2)+G_{2'\downarrow}(d,t_1-t_2)\bar G_{1\downarrow}(d,t_1-t_2)\right).
\end{multline}

\subsection{Collider configuration}

\subsubsection{Zero temperature noise}

Here we only present the result at zero temperature, and the finite-temperature result can be obtained in a similar way. 
The finite temperature version used to calculate numerically the cross-correlations shown in Figs.~\textbf{3} and \textbf{4} in the Main text and Supplementary Figure~\ref{ED_BlobAllV} is provided in the subsection `Finite temperature expressions'. 

Integrating over $\delta\tau$, the $T=0$ correlators given by Eqs.~(\ref{current_1_mu}) and (\ref{current_2_mu}) yield the cross-correlation noise at zero frequency as a function of the interacting Green functions $G_{1\downarrow}$ and  $G_{2'\downarrow}$:
\begin{multline}
S_{1\downarrow 2'\downarrow}(V)=2\frac{e^2}{2\pi\hbar}R_S T_S (R_1 T_1 + R_2 T_2) |eV| \\- 2 e^2 R_S T_S\tilde v^2\int_{-\infty}^{\infty} dt\left(G_{1\downarrow}(d,t)\bar G_{2'\downarrow}(d,t)+G_{2'\downarrow}(d,t)\bar G_{1\downarrow}(d,t)-(\mathrm{same}\ \mathrm{at}\ V=0 )\right),
\label{s12_full}
\end{multline}
where $\tilde v=\sqrt{v_{+}v_{-}}$.
Note that the noise $S_{1\downarrow 2'\downarrow}$ from Eq.~(\ref{s12_full}) can be written as a sum of the non-interacting contribution $S^{(0)}_{1\downarrow 2'\downarrow}$ given in Eq.~(5) and an interacting contribution $S^{\rm{(int)}}_{1_\downarrow 2'_\downarrow}$:
\begin{equation}
S_{1\downarrow 2'\downarrow}(V)=S^{(0)}_{1\downarrow 2'\downarrow}(V)+S^{\rm{(int)}}_{1\downarrow 2'\downarrow}(V),
\end{equation}
where 
\begin{equation}
S^{\rm{(int)}}_{1\downarrow 2'\downarrow}(V) =2 \frac{e^2}{2\pi\hbar} R_S T_S (R_1 T_2 + T_1 R_2) |eV|- 2 e^2 R_S T_S\tilde v^2\int_{-\infty}^{\infty} dt\left(G_{1\downarrow}(d,t)\bar G_{2'\downarrow}(d,t)+G_{2'\downarrow}(d,t)\bar G_{1\downarrow}(d,t)\right).
\end{equation}

The interacting Green functions can be obtained from refermionization, for example we have
\begin{equation}
\bar G_{1\downarrow}(d,t)=-\frac{i}{2\pi t \tilde{v}} \exp(-\tfrac{i}{2\hbar}eVt) \bar K_{1\downarrow}(d,t),
\end{equation}
where the function $\bar K_{1\downarrow}(d,t)$ is defined as
\begin{equation}\label{eqK1down}
\bar K_{1\downarrow}(d,t)=\frac{\left\langle \mathcal{S}^{\dagger}_1 \exp\left(-i\pi \mathcal N_{A_+}(d,t)+i\pi\mathcal{N}_{A_-}(d,t)\right)\mathcal{S}_1\right\rangle}{\left\langle \exp\left(-i\pi \mathcal N_{A_+}(d,t)+i\pi\mathcal{N}_{A_-}(d,t)\right)\right\rangle}.
\end{equation}
Here, the averages are taken with respect to the filled Fermi seas at the corresponding chemical potentials, and the scattering matrix $\mathcal S_1$ corresponds to the rotation of fermions $\hat{\Psi}_{A_+},\hat{\Psi}_{A_-}$ by QPC$_1$ in the standard way (see also Eq.~(\ref{eq_scattering}) for QPC$_S$).
We have defined the particle number operators \cite{kovrizhin2012,kovrizhin2013} $\mathcal N_{A_\pm}(d,t)$ for the refermionized channels with velocities $v_{\pm}$ as 
\begin{equation}
\mathcal{N}_{A_{\pm}}(d,t)=\int_{-d/v_{\pm}}^{t-d/v_{\pm}}\ \hat{\Psi}_{A_{\pm}}^{\dagger}(0,\tau)\hat{\Psi}_{A_{\pm}}(0,\tau)d\tau
\end{equation}
where the fermion operators are given in the interaction representation with tunneling at QPC$_1$ treated as interaction.
These particle number operators count the number of particles passing position $x=0$ in a time window $(-d/v_{\pm},-d/v_{\pm}+t)$. 

We note that functions $K$ have a form similar to the full counting statistics (FCS). At large distances the two exponents are uncorrelated, and these functions can be analysed by the methods developed for the FCS. 
At intermediate distances we have to rely on calculations using fermionic determinants \cite{kovrizhin2013}. 
In order to numerically calculate their values, we first obtained analytically the matrix elements of the FCS exponents with respect to the filled Fermi seas in the corresponding channels with one extra particle/hole.
Then, using these matrix elements in the expressions given in terms of fermionic determinants, we calculated the functions $K(d,t)$ (details will be provided in D.~Kovrizhin in preparation, see also Appendix A in Ref.~\citenum{kovrizhin2013}).

Similarly, the Green function for channel $2'_\downarrow$ is obtained as
\begin{equation}
G_{2'\downarrow}(d,t)=-\frac{i}{2\pi t \tilde{v}}\exp(\tfrac{i}{2\hbar}eVt) K_{2'\downarrow}(d,t),
\end{equation}
where
\begin{equation}
K_{2'\downarrow}(d,t)=\frac{\left\langle \mathcal{S}^{\dagger}_2 \exp\left(-i\pi \mathcal N_{A'_+}(d,t)+i\pi\mathcal{N}_{A'_-}(d,t)\right)\mathcal{S}_2\right\rangle}{\left\langle \exp\left(-i\pi \mathcal N_{A'_+}(d,t)+i\pi\mathcal{N}_{A'_-}(d,t)\right)\right\rangle}.
\end{equation}

The results at $T=0$ could be further simplified in the limit of large distances $d\to\infty$, where the exponents decouple (so they are independent of the distance $d$), and we consider the symmetric case $T_1=T_2$ where $S^{\rm{(0)}}_{1\downarrow 2'\downarrow}=0$. 
We have in this limit
\begin{equation}\label{S_inf}
S_{1\downarrow 2'\downarrow}^{d\to\infty}(V)=S^{\rm{(int)d\to\infty}}_{1\downarrow 2'\downarrow}(V) =+4 \frac{e^2}{2\pi\hbar} R_S T_S \left(R_1 T_1 |eV|+ \frac{\hbar}{2\pi} \times\int_{-\infty}^{\infty} \frac{dt}{t^2}\left(\left|\left\langle e^{-i\pi \mathcal N_{A_+}(t)}\right\rangle\left\langle e^{i\pi \mathcal N_{A_-}(t)}\right\rangle\right|^2-1\right)\right),
\end{equation}
where the averages are taken with respect to the non-interacting Fermi seas transformed by the scattering matrix $\mathcal{S}_{1,2}$. 
Note first that in the case of $T_1=T_2=1/2$ one can observe numerically that the averages are well-described by the analytical expression (see also the Supplementary Material of Ref.~\citenum{kovrizhin2012}):
\begin{equation}
\left|\left\langle e^{i\pi \mathcal N_{A_{\pm}}(t)}\right\rangle\right|^2=e^{-(eV t/4\hbar)^2}.
\end{equation}
Using this expression (which seems to be exact, but we do not have a proof), one can write the noise in the following form
\begin{equation}
S_{1\downarrow 2'\downarrow}^{d\to\infty}(V,T_1 = T_2 = 1/2) =\frac{e^2}{2\pi\hbar} R_S T_S |eV| \left(1-\sqrt{\frac{2}{\pi}}\right).
\end{equation}
We note that the cross-correlations are positive, compared to zero cross-correlations in the case of $T_1=T_2$ in the non-interacting limit.

\subsubsection{Dilute beam asymptotics}
In the case of dilute beams with equal transmission $T_1=T_2=T\rightarrow \{0,1\}$  and at $d\to\infty$ we can use the asymptotics developed in the theory of FCS. At short times the product of the exponents in the Eq.~(\ref{S_inf}) has a quadratic behaviour $1-\alpha t^2$, where constant $\alpha$ depends on the tunneling, so the integral converges at short times. This contribution is trivial, and below we focus on the long-time asymptotics. 

The long-time asymptotics can be calculated using the Fisher-Hartwig approach for the Toeplitz matrices developed in Ref.~\citenum{Gutman_2011}. 
Note that in the Supplemental Material of Ref.~\citenum{kovrizhin2012} regarding the equilibration of edge states, we showed a comparison of these asymptotics with the numerical results, and it was pointed out that the asymptotics break down at tunneling T=1/2 (see above).

Let us first reproduce the equations obtained in Ref.~\citenum{Gutman_2011}. 
We consider a double-step Fermi distribution at zero temperature
\begin{equation}
n(\varepsilon)=T\ n_0(\varepsilon-\mu_1)+R\ n_0(\varepsilon-\mu_2),
\end{equation}
where $n_0(\varepsilon)=\theta(-\varepsilon)$ is a step function, $\mu_2>\mu_1$ are the chemical potentials, $R=1-T$ is the reflection probability of the QPC, and $V=\mu_2-\mu_1$ is bias voltage.

We further introduce the following constants,  taking $\delta$ the fractionalization parameter in $\langle e^{-i\delta \hat N(t)}\rangle$ to be $\delta=\pi$ and assuming \cite{due_to}
$R<1/2$ 
\begin{equation}
\beta_1 =-\frac{i}{2\pi}\ln(1-2 R),\ \ \ \beta_0 =-\frac{1}{2}-\beta_1,
\end{equation}
as well as the dephasing time
\begin{equation}
t_\phi^{-1}=-\frac{eV}{2\pi\hbar}\ln(1-4 R T).
\end{equation}
With these definitions, the asymptotics of $\left\langle e^{-i\pi \hat N(t)}\right\rangle$ at long times (normalised to the equilibrium value), are given by the following expression obtained in Ref.~\citenum{Gutman_2011}
\begin{equation}
\Delta(t)=\left\langle e^{-i\pi \hat N(t)}\right\rangle_{\rm{norm}}\sim e^{-t/2 t_\phi}(V t)^{\frac{1}{4}-\beta_0^2-\beta_1^2}=(1-4 R T)^{eV t/4\pi\hbar}(V t)^{\ln^2(1-2 R)/2\pi^2}(V t)^{i \ln(1-2 R)/2\pi}.
\end{equation}

\noindent
Now let us use these expressions to calculate the noise. We need the absolute value of $\Delta(t)$, which reads
\begin{equation}
|\Delta(t)|\sim (1-4 R T)^{eV t/4\pi\hbar}(V t)^{\ln^2(1-2 R)/2\pi^2}.
\end{equation}
This function has an exponential decay times a power-law, and for small reflections $R\ll1$ we get
\begin{equation}
|\Delta(t)|\sim (1-4 R)^{eV t/4\pi\hbar}(V t)^{2 (R/\pi)^2}.
\end{equation}
In order to obtain the tunneling-dependent contribution to the noise we need to integrate these asymptotics
\begin{equation}
\tilde s(V,R)=2\int_\tau^{\infty}\frac{1}{t^2}(|\Delta(t)|^4-1)dt,
\end{equation}
where we have introduced a cutoff $\tau$. 
Note that if we do not assume the smallness of the reflection $R$, then we have to calculate the following integral:
\begin{equation}
\tilde{s}(V,R)=2\int_\tau^{\infty}dt\ \frac{1}{t^2} \left((1-4 R T)^{eV t/\pi\hbar}(eV t/\hbar)^{2\ln^2(1-2 R)/\pi^2}-1\right).
\end{equation}
One can express this integral in terms of a special function (where we introduce a dimensionless cutoff $\Theta=eV\tau/\hbar$), but we only need its asymptotic development at small $R$, which gives
\begin{multline}
S_{1\downarrow2'\downarrow}(V,T_1=T_2)/(4R |e V|R_S T_S e^2/2\pi\hbar)\simeq\frac{4}{\pi^2}\ln(R)+1-\frac{4}{\pi^2}(\ln(4\Theta/\pi)+\gamma-1)\\
+\frac{4}{\pi^2}R\ln(R)+R\left(-1+\frac{8}{\pi^3}(1-\Theta)+\frac{4}{\pi^2}(\ln(4\Theta/\pi)+\gamma)\right),
\end{multline}
which has the most singular terms containing logarithms independent of the cutoff, as well as the constant term and the term proportional to $R$,
which do depend on the cutoff. Using the structure of this expansion with respect to $R$, we can obtain the numerical values of the cutoff-dependent terms by fitting the numerical results obtained from the theory. This gives the following terms in the noise asymptotics at small $R$
\begin{equation}
S_{1\downarrow 2'\downarrow}(V,R)/(4 R |e V|R_S T_S e^2/2\pi\hbar)\simeq \frac{4}{\pi^2} \ln(R)+\frac{4}{\pi^2} R\ln(R)-0.2995 R+0.9427.
\label{eq_asspt}
\end{equation}

Finally, we can rewrite this in the notation used in the Main text in terms of the generalized Fano factor $P$ and using the symmetry $\tau \rightarrow 1-\tau$:

\begin{equation}\label{cross-correlation-EMP}
P \simeq  \frac{4}{\pi^2}\ln \tau_\mathrm{s} +  \frac{4}{\pi^2}\tau_\mathrm{s} \ln \tau_\mathrm{s} - 0.2995 \tau_\mathrm{s} + 0.9427.
\end{equation}

\noindent 
For $\tau_\mathrm{s}=0.05$ we obtain $P \simeq -0.35$, close to the experimental value in Fig.~\textbf{3} of the Main text. 

\subsubsection{Finite temperature expressions}
Here, we provide the finite temperature expression for the cross-correlation noise at zero frequency in the standard collider configuration shown Supplementary Fig.~\ref{fig:collider}. 
The following expression, obtained along the same lines described above for $T=0$, was computed numerically for the data-theory comparison:
\begin{multline}\label{eq:ScollToEC}
S_{1\downarrow 2'\downarrow}(V)=2\frac{e^2}{2\pi\hbar}e V R_S T_S (R_1 T_2+T_1 R_2)  (\coth(e V/2 k_B T)-2 k_B T/e V)\\
+4\frac{e^2}{2\pi\hbar}e V R_S T_S (\pi k_B T/e V)^2\ \mathrm{Re}\int_{-\infty}^{\infty}\frac{d\tau}{2\pi\hbar}\ \sinh^{-2}(\pi k_B T\tau/\hbar)(K_{1\downarrow}(d,\tau) \bar K_{2'\downarrow}(d,\tau)-1).
\end{multline}

We also provide the finite temperature expression for the cross-correlation noise at zero frequency in the alternative configuration where the analyzer QPC$_S$ is tuned to couple the two inner edge channels ($2'_\uparrow$ and $1_\uparrow$, which are not directly excited by the source QPC$_{1,2}$) with the transmission probability $T_S$:
\begin{equation}\label{eq:ScollTiEC}
S_{1\uparrow 2'\uparrow}(V)=
+4\frac{e^2}{2\pi\hbar}e V R_S T_S (\pi k_B T/e V)^2\ \mathrm{Re}\int_{-\infty}^{\infty}\frac{d\tau}{2\pi\hbar}\ \sinh^{-2}(\pi k_B T\tau/\hbar)(K_{1\uparrow}(d,\tau) \bar K_{2'\uparrow}(d,\tau)-1).
\end{equation}

\subsection{Electron energy distribution configuration}

The same refermionization approach can be used to address the configuration implemented to observe the electron energy distribution. 
The different biasing corresponds to applying a second bias voltage $V_\mathrm{p}$ to both the edges 1 and 2 (following the labels in Supplementary Fig.~\ref{fig:collider}), which changes the chemical potentials provided by Eq.~(\ref{muASpm}) for the collider configuration.
In that case, QPC$_2$ does not play any role and the edges 1 and 2 can be merged into a single edge, as schematically pictured in Supplementary Fig.~\ref{fig:distri}.

\begin{figure*}[!htb]
\renewcommand{\figurename}{\textbf{Supplementary Figure}}
\renewcommand{\thefigure}{\textbf{\arabic{figure}}}
	\centering
	\includegraphics[width=12cm]{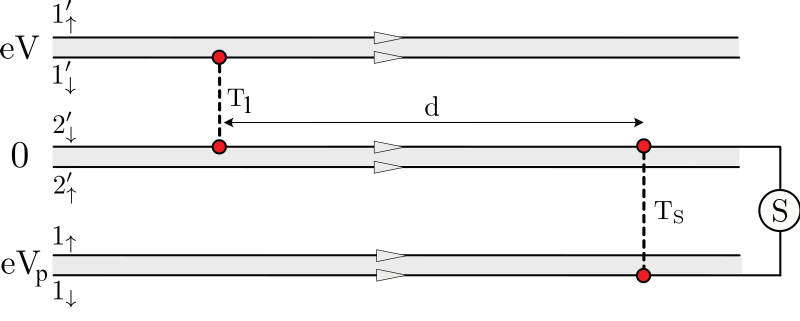}
	\caption{
	\footnotesize
         Schematic of the configuration pertaining to the distribution measurement. 
	 The channels $1'_\downarrow$ and $2'_\downarrow$ are coupled via a source QPC positioned at $x=0$ with the tunneling probability $T_1$.  The channels $2'_\downarrow$ and $1_\downarrow$ are coupled by the analyzer QPC with tunneling probability $T_S$ positioned at $x=d$, after which the current correlations are measured. 
 }	\label{fig:distri}
\end{figure*}

Along the same lines as in the collider configuration, the cross-correlations at zero-frequency and finite temperature $T$ in the electron energy distribution configuration can be written:
\begin{multline}\label{eq:S12distrib}
S_{1\downarrow 2'\downarrow}(V, V_\mathrm{p})=4\frac{e^2}{2\pi\hbar}e V R_S T_S\times \Biggl(\frac{k_B T}{e V}-\frac{V_\mathrm{p}}{2 V}\tanh^{-1}\left(\frac{e V_\mathrm{p}}{2k_B T}\right)\\+\frac{e V}{2}\left(\frac{\pi k_B T}{e V}\right)^2\mathrm{Re}\int_{-\infty}^{\infty}\frac{d\tau}{2\pi\hbar}\  e^{i e V_\mathrm{p} \tau/\hbar}\ \mathrm{sinh}^{-2}(\pi k_\mathrm{B} T \tau/\hbar)(\bar K_{2'\downarrow}(d,\tau)-1) \Biggr),
\end{multline}
where the channels $A'_{+}$ and $A'_{-}$ are set to voltage $V$ and $0$, respectively and a similar expression for the measurement in channel $2'\uparrow$ with the function $K_{2'\uparrow}(d,t)$. 
The numerical calculations of $S_{12}$ based on Eq.~(\ref{eq:S12distrib}) are compared to the data in Fig.~\textbf{2} in the Main text, as well as in Supplementary Figs.~\ref{ED_Distributions},\ref{ED_LoWTauDist},\ref{ED_Inner}.

\vspace{8mm}
\section{Anyonic exchange theory}\label{sec:anyons}

The left and right source QPC inject electrons towards the central analyzer where cross-correlations are measured. Because of interchannel interaction, the electron wave-packets of charge $e$ separate into twin wave-packets carrying each half of the electron charge $e/2$. They correspond to the neutral and charge modes of the two copropagating channels progressing with distinct velocities $v_\mathrm{n}$ and $v_\mathrm{c}>v_\mathrm{n}$. 
The resulting beam has a mixed nature: the injection is random (and poissonian for $\tau_\mathrm{s} \ll 1$) for the center of mass of the twin wave-packets, whereas the distance between them is purely deterministic. 

An alternative way of producing fractional charges in the integer quantum Hall case was proposed in Ref.~\citenum{Morel_IntegerCollider_2022} by using a metallic quantum dot (QD) as a source. Only the neutral mode is then excited and the train of charge $e/2$ is predicted to be entirely randomly distributed. Using a fully quantum bosonization approach, the cross-correlations out of an analyzer QPC were computed~\cite{Morel_IntegerCollider_2022} to be 
 \begin{equation}\label{cross-correlation-QD}
S_{12} \sim   \tau_\mathrm{c} \, S_\Sigma^{QD} \,\frac{2 \left( \sin{\theta} \right)^2}{\theta^2}\ln{\tau_\mathrm{s}}
\end{equation}
 in the balanced case with two sources, $\tau_\mathrm{s} \ll 1$, and with $\tau_\mathrm{c}$ the analyzer transmission. 
 $\theta = \pi /2$ is the mutual exchange phase between an electron and a fractional charge $e/2$. 
 This results from the partition noise of the analyzer by electron-hole pair creation yielding a negative contribution. Remarkably, the partition term was identified~\cite{Morel_IntegerCollider_2022,mora2022} to be the consequence of a braiding mechanism in 1+1 (space and time) dimension involving the fractional excitation encircling the path of the electron-hole pair. 

 At low temperature, the noise of the sources is mostly shot noise, proportional to the granular charge of the signal. This is $e$ for our geometry as electrons tunnel from the source QPC whereas it is $e/2$ with the metallic quantum dot of Ref.~\citenum{Morel_IntegerCollider_2022}. Therefore, the incoming shot noise is twice larger in our experiment, $S_\Sigma = 2 S_\Sigma^{QD}$, for the same quasiparticle current. Normalizing the cross-correlation of Eq.~\eqref{cross-correlation-QD} with $\tau_\mathrm{c} S_\Sigma $ (see Eq.~(2) in the Main text), we find for the partition noise $P = (4/\pi^2) \ln \tau_s$, exactly as the leading term in 
 Eq.~\eqref{cross-correlation-EMP} derived within the complete non-perturbative theory applied to our geometry. We can draw two consequences from this result:
 \begin{enumerate}
     \item the non-perturbative theory is in excellent agreement with our experimental data. Although the braiding mechanism is not transparent in this theory, the exact asymptotic matching in the dilute limit $\tau_\mathrm{s} \ll 1$ with the braiding theory of Ref.~\citenum{Morel_IntegerCollider_2022} shows that this theory encompasses the braiding process. The partition noise is the same in Ref.~\citenum{Morel_IntegerCollider_2022} and in our experiment.
     \item our experiment nonetheless differs from the quantum dot fractionalization proposal in that the noise generated by the sources is two times larger. We thus need to multiply the first term of Eq.~\eqref{cross-correlation-QD} by a factor $2$ and obtain
      \begin{equation}\label{cross-correlation-our}
          P = \frac{S_{12} }{\tau_\mathrm{c} S_\Sigma} \sim \frac{\left( \sin{\theta} \right)^2}{\theta^2}\ln{\tau_\mathrm{s}} =  \frac{4}{\pi^2} \ln{\tau_\mathrm{s}}.
      \end{equation}
 \end{enumerate}

\vspace{8mm} 
\section{Random or deterministic injected signal} 

\begin{figure*}[!htb]
\renewcommand{\figurename}{\textbf{Supplementary Figure}}
\renewcommand{\thefigure}{\textbf{\arabic{figure}}}
	\centering
	\includegraphics[width=12cm]{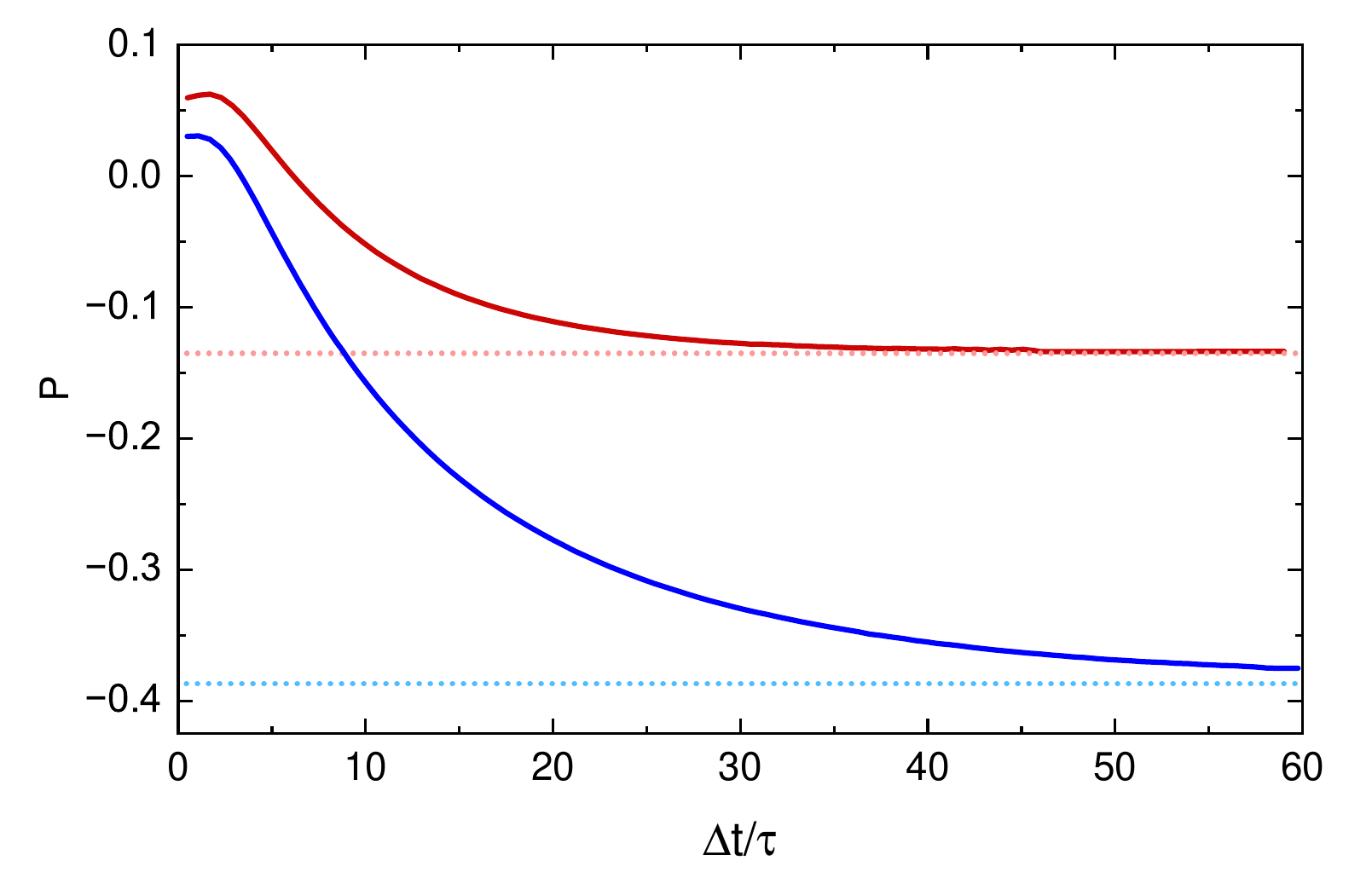}
	\caption{
	\footnotesize
	Cross-correlations P  (plain lines) as function of the time delay $\Delta t$ between the twin pulses.  The dotted lines give $P$ for random $e/2$ pulses. $\tau/\tau_s = 0.005$ and  $\tau/\tau_s = 0.01$ are shown in red and blue.
 }	\label{fig1}
\end{figure*}

We turn here to the issue of random vs deterministic for the injected beam since the distance between twin pulses is fixed. The asymptotic matching of the non-perturbative theory with the theory of Ref.~\citenum{Morel_IntegerCollider_2022}, where the injection is random, suggests that the distinction is not important as long as the time splitting between twin wave-packets exceeds largely the mean distance between subsequent pairs. Ref.~\citenum{mora2022} has shown that the injection by the source quantum point contacts, as far as cross-correlations are concerned, is generally equivalent to the driving by an ohmic contact with a series of random quantized voltage pulses. We thus consider here a series of twin pulses, with time splitting $\Delta t$, sent with a Poissonian distribution. Following Ref.~\citenum{mora2022}, we use the Kubo formula and obtain for the renormalized cross-correlation noise
\begin{equation}
P= - \frac{1}{4 \pi^2 } \int_{-\infty}^{+\infty} d t \frac{e^{- \frac{\tau_s}{\tau} \left( g (t) + g^*(t) \right) }}{(0^+ + i t)^2},
\end{equation} 
where $\tau$ is the temporal width of individual pulses. We have introduced the function
\begin{equation}\label{functiong2}
    g(t) = \int_{-\infty}^{+\infty} d t' \, \left( 1 - e^{i \left( \phi_0 (t-t') - \phi_0 (-t') \right)}  \right),
\end{equation}
where, for twin Lorentzian pulses of charge $e/2$,
 \begin{equation}
   \phi_0 (t) = \frac{e}{\hbar} \int^t_{-\infty} d t' \, V_0 (t') \qquad \qquad  V_0 (t) = \frac{h}{2 e} \left(  \frac{\tau/\pi}{(t-\Delta t/2)^2 + \tau^2}   +\frac{\tau/\pi}{(t+\Delta t/2)^2 + \tau^2}  \right).
 \end{equation}
Supplementary Fig.~\ref{fig1} illustrates the comparison of this calculation with a direct evaluation of $P$ for purely random pulses (and not series of twin pulses) of charges $e/2$, confirming that they coincide in the limit of very long time delay $\Delta t \gg \tau/\tau_\mathrm{s}$.

\vspace{8mm}

\section{Phenomenological fermionic theory}

\hspace{\parindent} Here we consider an alternative model which also yields negative cross-correlations in the dilute limit. It is a phenomenological fermionic model which takes interactions into account via energy redistribution into hot Fermi distribution functions. 

In the collider geometry discussed in the Main text, we consider a channel which starts out with a double-step initial distribution, and following inter-channel energy transfer it has  the relaxed Fermi distribution with an effective temperature 

\begin{equation}
    T^* = \sqrt{T^2+\frac{3}{2}\tau_\mathrm{s}(1-\tau_\mathrm{s})\left(\frac{eV_\mathrm{s}}{\pi k_\mathrm{B}}\right)^2}
    \label{eq:Inter_Texc},
\end{equation}

\noindent where $V_\mathrm{s}$ is bias, $T$ is base temperature, and $\tau_\mathrm{s}$ is the transmission of the source QPC \cite{LeSueur_EnergyRelax_2010, Degiovanni_Plasmons_2010}. 

This kind of phenomenological description has previously been used for explaining experimental results (see, e.g., Ref.~\citenum{Ota_CrossPosNegNu=1_2017} at $\nu=1$ and Refs.~\citenum{Altimiras_SpectroscopieIQHE_2009, LeSueur_EnergyRelax_2010, Degiovanni_Plasmons_2010} at $\nu=2$). The cross-correlations have the form:

\begin{equation}
    S_\mathrm{12} = \tau_\mathrm{c}(1-\tau_\mathrm{c})S_\mathrm{\Sigma}-2\frac{e^2}{h}\tau_\mathrm{c}(1-\tau_\mathrm{c})\int_{-\infty}^{\infty} \left( f_\mathrm{L}(1-f_\mathrm{R}) + f_\mathrm{R}(1-f_\mathrm{L}) \right) d\varepsilon,
    \label{eq:Inter_CrossButtiker1}
\end{equation}

\noindent with $L$ ($R$) denoting the left (right) distribution in the outer channel and $S_\mathrm{\Sigma}$ the total noise incoming from both sources which is equal to $2\times2\frac{e^2}{h}\tau_\mathrm{s}(1-\tau_\mathrm{s})eV_\mathrm{s}\left(\coth{\frac{eV_\mathrm{s}}{2k_\mathrm{B}T}}-\frac{2k_\mathrm{B}T}{eV_\mathrm{s}}\right)$ in the case of balanced beams. Plugging the Fermi distribution with temperature from  Eq.~(\ref{eq:Inter_Texc}) into Eq.~(\ref{eq:Inter_CrossButtiker1}), we find

\begin{equation}
    S_\mathrm{12} = \tau_\mathrm{c}(1-\tau_\mathrm{c})S_\mathrm{\Sigma}-4\frac{e^2}{h}\tau_\mathrm{c}(1-\tau_\mathrm{c})k_\mathrm{B}T^*,
    \label{eq:Inter_CrossButtiker1bis}
\end{equation}

 \noindent which at $T = 0$ yields the Fano factor 

\begin{equation}
    P = 1-\frac{\sqrt{3}}{\pi\sqrt{2\tau_\mathrm{s}(1-\tau_\mathrm{s})}}.
    \label{eq:Inter_PButtiker}
\end{equation}


\begin{figure}[!htb]
\renewcommand{\figurename}{\textbf{Supplementary Figure}}
\renewcommand{\thefigure}{\textbf{\arabic{figure}}}
    \centering
    \includegraphics[width=10cm]{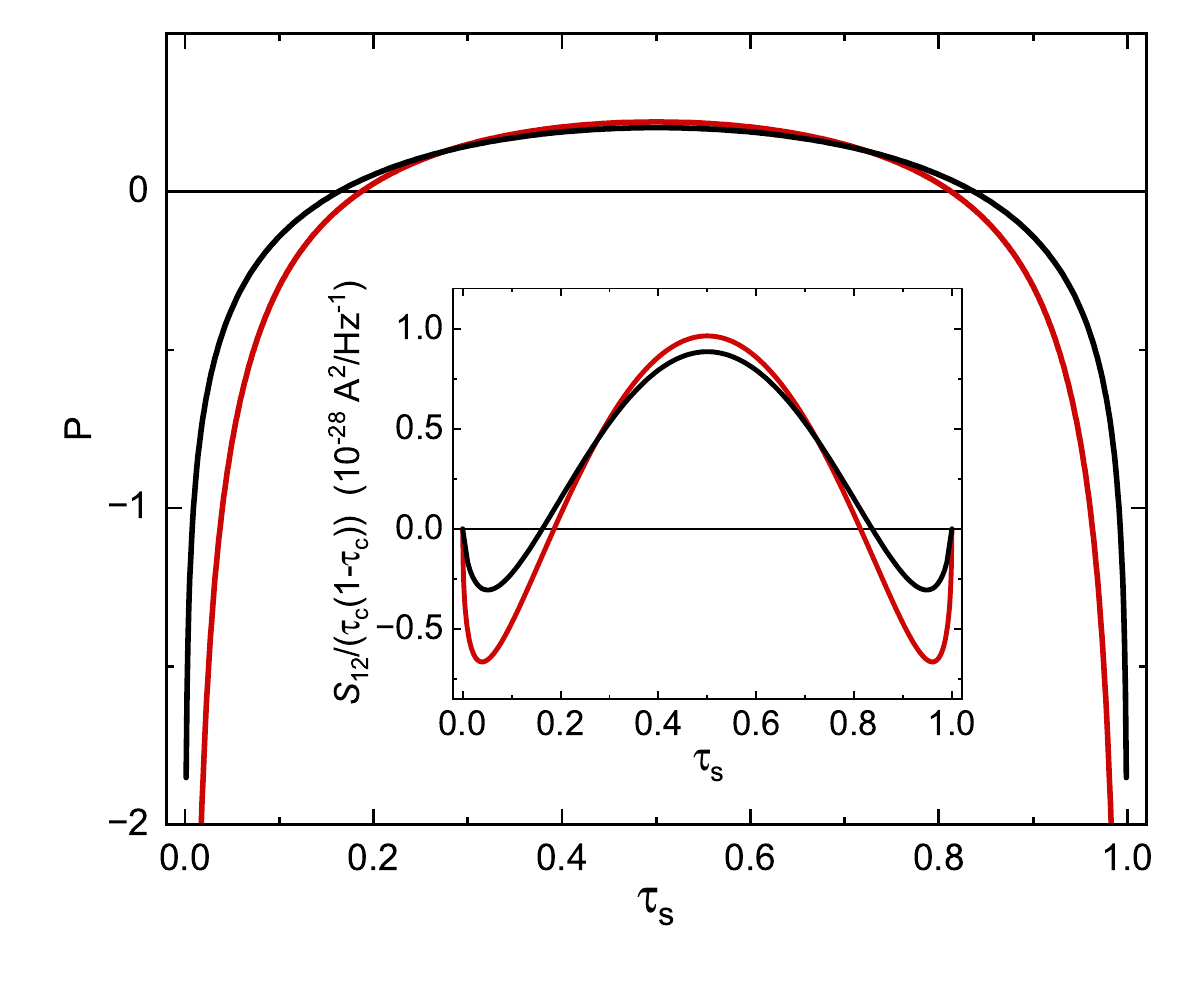}
    \caption[Cross-correlations and Fano factor in the phenomenological vs the non-perturbative framework.]{\textbf{Comparison between the phenomenological and the non-perturbative framework.}
    Inset: Renormalized cross-correlations $S_{12}/(\tau_\mathrm{c}(1-\tau_\mathrm{c}))$ versus the source transmission $\tau_\mathrm{s}$ are shown at $T=$~\SI{0}{K}. The black line is the prediction of the non-perturbative theory, and the red line is the prediction of the phenomenological fermionic theory. Main panel: Fano factor $P$ calculated in the two approaches. 
    }
    \label{fig:Inter_Blob_ThyFermi}
\end{figure}

\medskip

Therefore, this alternative description of the system with interacting fermions can also lead to non-zero cross-correlations with the same change of sign between the dilute regime and the $\tau_\mathrm{s}\sim0.5$ regime, due to the relative importance of the positive source noise redistribution and negative cross-correlation contribution.  Moreover, as seen in Supplementary Fig.~\ref{fig:Inter_Blob_ThyFermi}, the shape of the prediction for both models is similar, although the quantitative values are different. We therefore conclude that the mere presence of negative cross-correlations cannot be directly attributed to the anyonic exchange phase, since a simple fermion model also qualitatively predicts it. In order to be able to attribute the negative signal to the fractional statistics of the involved charges, we need to validate the non-perturbative approach by complementary distribution measurements, as we have done in the Main text.
Note that we cannot directly rule out the phenomenological fermionic theory because this would require one to compare the observed distributions with specific predictions. 
However this phenomenological theory does not allow one to make specific predictions regarding the evolution of the distribution as this would require to introduce a choice for the rate of inelastic collisions as a function of the exchanged energy.

\vspace{8mm}
\section{Fitting procedure to determine $\delta t$}
\noindent

We outline the procedure used to obtain the only fitting parameter of the theory, the time delay $\delta t$ between the arrival of fractionalized $e$/2 charges at the analyzer QPC, namely $\delta t=d/v_\mathrm{n}-d/v_\mathrm{c}$, with $d$ the distance between source and analyzer quantum point contacts and $v_\mathrm{c,n}$ the velocities of charged and neutral mode.

In the source-analyzer configuration we measure the cross-correlations $S_{12}(V_\mathrm{p})$ which yield the distributions $f(\varepsilon = eV_\mathrm{p})$ (Main text Eq.~(1)). 
Cross-correlations $S_{12}(V_\mathrm{p})$ contain a big contribution from the equilibrium noise :
\begin{equation}
S_\mathrm{12}^{ \;\;\; 0} = 2 \frac{ e^2}{h} \tau_\mathrm{s}(1-\tau_\mathrm{s})|e V_\mathrm{p}|
\left(
\mathrm{coth}\left(\frac{eV_\mathrm{p}}{2k_{\mathrm{B}}T}\right)  - \frac{2k_{\mathrm{B}}T}{eV_\mathrm{p}} 
\right).
\end{equation}

\noindent
After subtraction we obtain $dS_\mathrm{12}(V_\mathrm{p}) = S_{12}(V_\mathrm{p}) - S_\mathrm{12}^{\;\;\; 0}(V_\mathrm{p})$, which is very sensitive to the value of $\delta t$. We find the best value by a least-squares method in the data subset corresponding to the [\SI{59}{\uV},\SI{82}{\uV}] range of bias voltage. We chose this range because it corresponds to the regime of fully fractionalized charge. 
As given in the Main text, we have determined $\delta t$ = \SI{64}{ps}. With $d = $ \SI{3.1}{\um} measured in the SEM photo (Fig.\textbf{1}c, Main text), we obtain $d/\delta t = $ \SI{5d4}{m.s^{-1}}.
We find slightly different values of $\delta t$ for left and right side, namely $\delta t_{\mathrm{L}} = 68 \pm 2$ ps and $\delta t_{\mathrm{R}} = 60 \pm 2$ ps. If we assume that the velocity difference between the fast and slow mode is the same on both sides, this yields the left and right distance of $3.3 \pm 0.1$\SI{}{\um} and $2.9\pm 0.1$\SI{}{\um}. This is plausible if we consider the SEM photo which reveals a slightly longer distance between the source and analyzer QPC on the right-hand side (Fig. \textbf{1}c, Main text). The difference can also originate from the screening details in the edge. However, since the theory is developed for equal lengths on the left and right, we adopt the mean value of $\delta t =$ \SI{64}{ps} which we use throughout.

\vspace{8mm}
\section{Distributions}

\subsection{Distributions at intermediate bias voltages}

In Supplementary Fig. \ref{ED_Distributions} we expand on the data shown in Fig. \textbf{2} of the Main text and show the full evolution of the distribution function for bias voltages between \SI{12}{\uV} and \SI{82}{\uV}. The injection and measurement take place on the outer channel. We see the relaxation from the double-step at lower bias into a single broader-step distribution in the range 47-70 \SI{}{\uV}. 

At $V_\mathrm{s} = $ \SI{82}{\uV} we have some inter-channel tunneling starting to take place, see Section XI below. 

\begin{figure*}[!htb]
\renewcommand{\figurename}{\textbf{Supplementary Figure}}
\renewcommand{\thefigure}{\textbf{\arabic{figure}}}
	\centering
	\includegraphics[width=17.9cm]{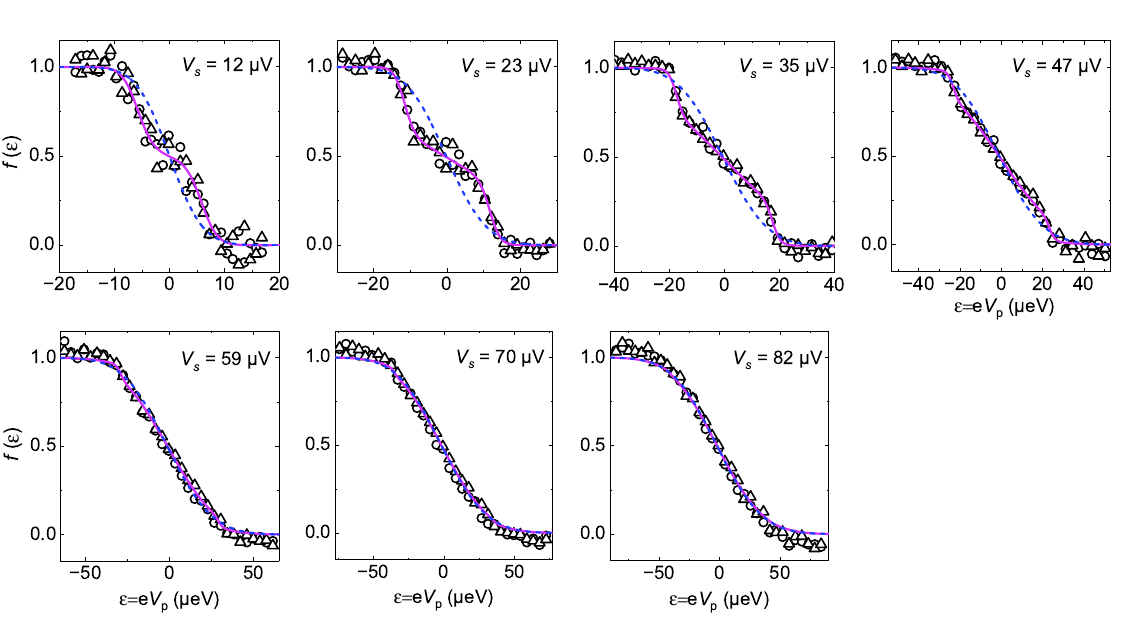}
	\caption{
	\footnotesize
	\textbf{Electron energy distribution spectroscopy at intermediate voltages.}
	 Electron energy distribution $f$ vs injected energy $\varepsilon$ for the source bias voltage $V_\mathrm{s}$ given in each panel, ranging from \SI{12}{\uV} to \SI{82}{\uV}.
   The additional $V_\mathrm{s}$ complete the three values shown in Fig.~\textbf{2} in the Main text.
   Squares and triangles correspond to $V_\mathrm{s}$ applied, respectively, to the left and right source QPC.
   Both source and analyzer transmissions are set to $0.5$.
   Continuous purple lines show the theoretical predictions for $\delta t=$ \SI{64}{ps} and the dashed blue lines those for $\delta t = \infty$. 
   }
    \label{ED_Distributions}
\end{figure*}


\subsection{Distributions at low transparency}

In order to verify that the charge fractionalization in the dilute limit does not deviate from the expected behavior, we measured the distributions at source transmission $\tau_\mathrm{s} = 0.05$ and $\tau_\mathrm{s} = 0.95$, shown in Supplementary Fig. \ref{ED_LoWTauDist}. In the inset we show $S_{12}$ used to obtain $f(\varepsilon)$ by derivation, see Eq.~(1) in the Main text. 

We see that the finite- and infinite-time predictions (purple and blue curve) are very close, and explain the data very well. With the same reasoning outlined in the Main text, from this we conclude that the full fractionalization has taken place for dilute beams. We shall use this result for the cross-correlation 'collider' measurement. 

\begin{figure*}[!htb]
\renewcommand{\figurename}{\textbf{Supplementary Figure}}
\renewcommand{\thefigure}{\textbf{\arabic{figure}}}
	\centering
	\includegraphics[width=17.0cm]{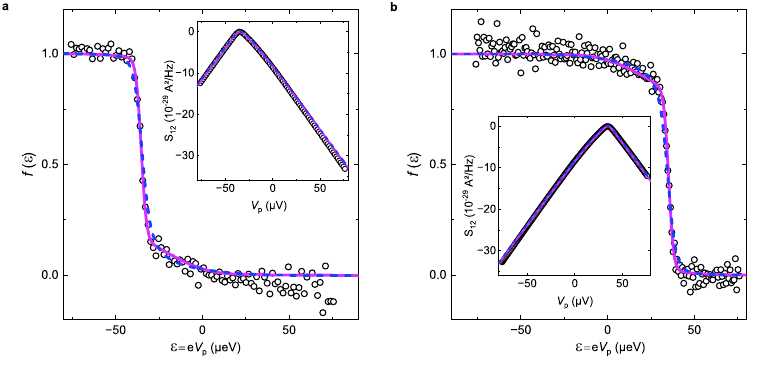}
	\caption{
	\footnotesize
  	\textbf{Electron energy distribution spectroscopy at high beam dilution.}  
 Energy distribution $f$ obtained from $S_{12}$ (insets) by derivation, see Main text, Eq.~(1). Panel \textbf{a} corresponds to $\tau_\mathrm{s}=$ 0.05 and panel \textbf{b} to $\tau_\mathrm{s}=$ 0.95.
 The analyzer transmission is set to $\tau_\mathrm{c} = 0.5$ and bias voltage to $V_\mathrm{s}=$ \SI{70}{\uV}. Injection is from the right QPC.
 Full purple and dashed  blue line are the predictions for $\delta t=$ \SI{64}{ps} and $\delta t = \infty$ respectively.   }
    \label{ED_LoWTauDist}
\end{figure*}

\vspace{5mm}
\subsection{Inner channel distributions}

\begin{figure*}[!htb]
\renewcommand{\figurename}{\textbf{Supplementary Figure}}
\renewcommand{\thefigure}{\textbf{\arabic{figure}}}
	\centering
        \includegraphics[width=17.9cm]{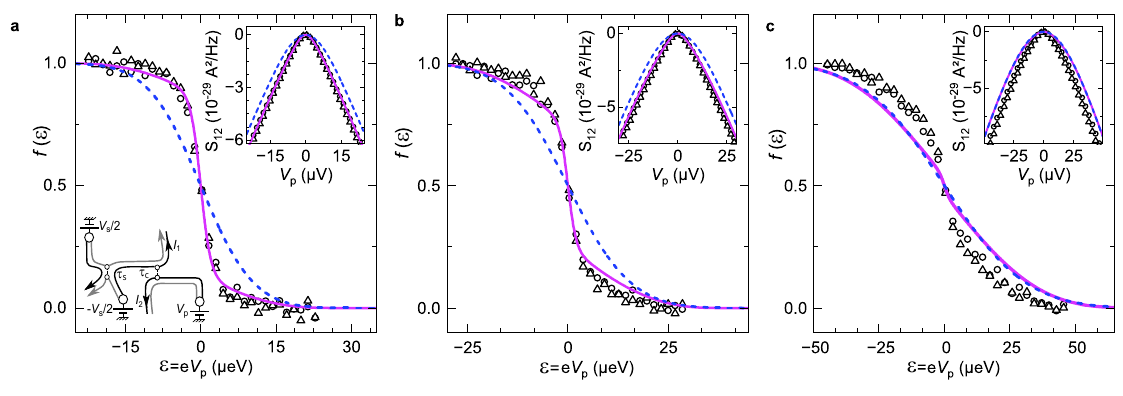}
	\caption{
        \footnotesize
        \textbf{Electron energy distribution spectroscopy along the inner edge channel.}
        \textbf{a,b,c} Energy distribution $f(\varepsilon)$ obtained by deriving the measured $S_\mathrm{12}$ (insets, see Eq.~(1) and plotted versus the probe energy $\varepsilon = e V_\mathrm{p}$, taken at bias voltage $V_\mathrm{s}=$ \SI{23}{\uV}, \SI{35}{\uV} and \SI{70}{\uV} from left to right. The injection is on the inner channel and the measurement on the outer (see schematic in \textbf{a}). Circles and triangles correspond respectively to the injection from the left and right source. Both source and analyzer transmissions are tuned to $\tau_\mathrm{s} = \tau_\mathrm{c} = 0.5$. Full purple and dashed blue line display the numerical predictions for $\delta t=$ \SI{64}{ps} and $\delta t = \infty$ respectively.
        }
    \label{ED_Inner}
\end{figure*}

\color{black}

In Supplementary Fig. \ref{ED_Inner} we show the measured distributions in the configuration where we inject into the inner channel, and measure on the outer, see the schematic in the inset. 
In this case we don't see the double-step at low bias, since there was no injection, but we see the broadening of the initial distribution when increasing bias. At full relaxation we expect the distribution on the injection channel to match that on the other channel. Theoretically this is the case (purple and blue curve compared to their counterpart in Supplementary Fig. \ref{ED_Distributions}), but experimentally we see that the relaxation is not complete. 


\section{Cross-correlations for a dilute vs non-dilute beam}
\noindent

\begin{figure*}[!htb]
\renewcommand{\figurename}{\textbf{Supplementary Figure}}
\renewcommand{\thefigure}{\textbf{\arabic{figure}}}
	\centering
	\includegraphics[width=18cm]{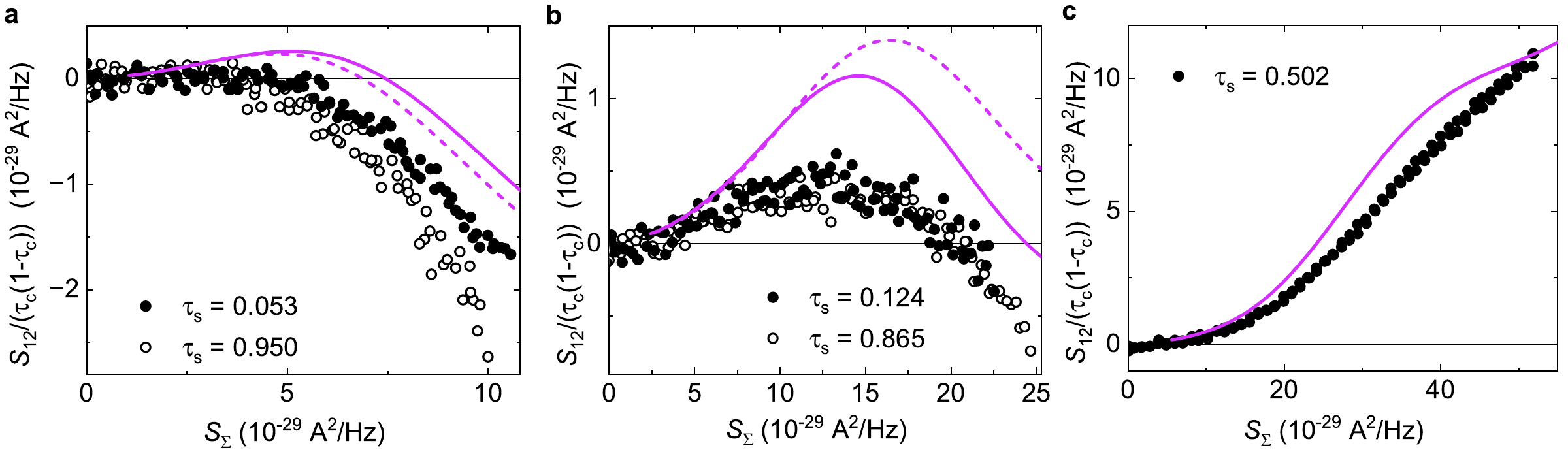}
	\caption{
	\footnotesize
	\textbf{}
	\textbf{a, b, c,} Cross-correlations $S_\mathrm{12}/(\tau_\mathrm{c}(1 - \tau_\mathrm{c}))$ vs source noise $S_\mathrm{\Sigma}$ at transparencies $\tau_\mathrm{s} = 0.053/0.95$ (\textbf{a}), $0.124/0.865$ (\textbf{b}) and $0.502$ (\textbf{c}). All quantum point contacts are tuned to outer channels. In panels \textbf{a} and \textbf{b} full circles denote data taken at low transparency $\tau_\mathrm{s}$, and open circles at transparency close to 1. The purple lines denote the prediction for $\delta t =$ \SI{64}{ps}, full lines for $\tau_\mathrm{s} = 0.053/0.124$ and dashed lines for $\tau_\mathrm{s} = 0.95/0.865$ in \textbf{a}/\textbf{b} respectively.
 	}	\label{Cross3tau}
\end{figure*}

In Supplementary Figure \ref{Cross3tau} we show cross-correlations $S_{12}/(\tau_\mathrm{c}(1-\tau_\mathrm{c}))$ vs $S_\mathrm{\Sigma}$ in three regimes when they are negative, at the transition, and positive. In Supplementary Fig.~\ref{Cross3tau}a we are interested in the low transparency $\tau_\mathrm{s} = 0.053$ (full circles) and its complement $\tau_\mathrm{s} = 0.95$ (open circles) (same data as in Main text Fig. \textbf{3}) .  In Supplementary Fig.~\ref{Cross3tau}b at $\tau_\mathrm{s} = 0.124, 0.865$ (full/open circles respectively) the system is in the intermediate regime where the cross-correlations change sign. In Supplementary Fig.~\ref{Cross3tau}c taken at $\tau_\mathrm{s} = 0.502$, the cross-correlations are positive in the full bias range.
The purple curves are the prediction of the non-perturbative theory for $\delta t$ = \SI{64}{ps}. Full curves are predictions for $\tau_\mathrm{s} = 0.053/0.124$ and dashed curves for $\tau_\mathrm{s} = 0.95/0.865$. There is a slight difference between the full and dashed curve in each panel because $\tau_\mathrm{s}$ and $1-\tau_\mathrm{s}$ are not exactly the same.
We see that the slopes are reasonably well reproduced by the theory, but the detailed agreement at low bias is absent. The prediction around zero bias is not available as it requires very long calculation times.

Supplementary Fig \ref{Cross3tau}a shows the same data as Fig. \textbf{3} in the Main text for $\tau_\mathrm{s} = 0.053/0.95$. The two datasets should yield a similar slope, i.e., a similar Fano factor. We notice that the experimental Fano factor is slightly different in the two cases. The slope of the negative part is $P \simeq -0.38$ at $\tau_\mathrm{s} \sim 0.05$, and $P \simeq -0.56$ at $\tau_\mathrm{s} \sim 0.95$. 
Some part of the discrepancy may be due to the tunneling between copropagating channels which starts to appear at large voltage (see Supplementary Figs. \ref{Tunneling} and \ref{Tunneling_dist} and the tunneling discussion below), 
or to the experimental particularities such as the nonequivalent paths towards the analyzer or the temperature difference between the ohmic contacts $\sim $ \SI{1}{mK} (see Methods). 
We also see asymmetry between $\tau$ and $1-\tau$ in Fig. \textbf{4} in the Main text, and in Supplementary Fig. \ref{ED_BlobAllV}. 

The uncertainties on measured source QPC transparencies are the following : 0.053 $\pm$ 0.002, 0.950 $\pm$ 0.001, 0.124 $\pm$ 0.004, 0.865 $\pm$ 0.003 and 0.502 $\pm$ 0.004. We are using the average of the high-bias region where the transmission dependence on bias is weak (cf. Fig \textbf{3}a in the Main text).

\section{Cross-correlations in the inner channel}
\noindent

\begin{figure*}[!htb]
\renewcommand{\figurename}{\textbf{Supplementary Figure}}
\renewcommand{\thefigure}{\textbf{\arabic{figure}}}
	\centering
	\includegraphics[width=12cm]{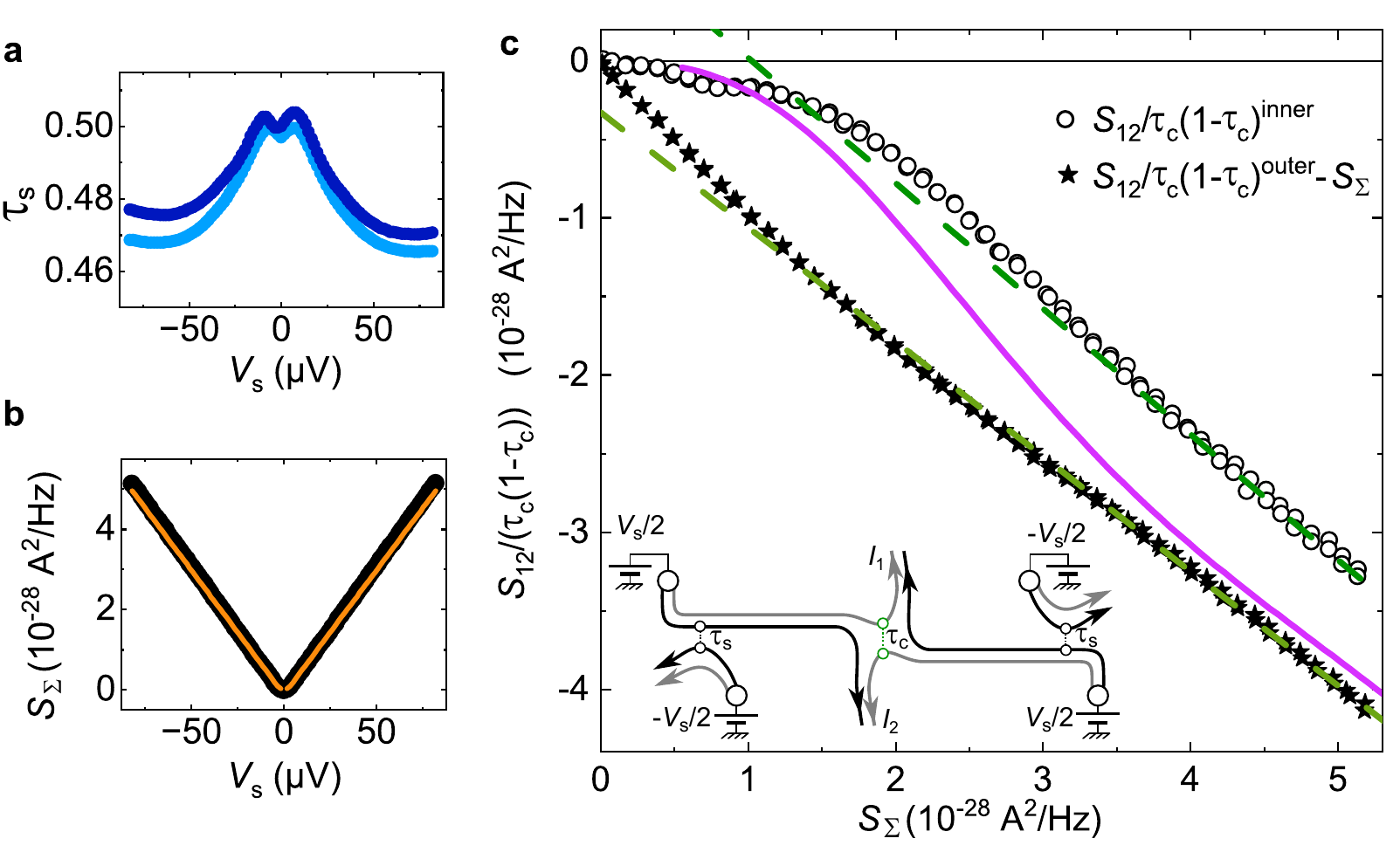}
	\caption{
	\footnotesize
    \textbf{Cross-correlations with the analyzer QPC probing the inner channel.}
    \textbf{a}, Measured left/right source QPC dc transmission as a function of bias voltage, shown in light/dark blue, respectively.
	 \textbf{b}, Source noise $S_\mathrm{\Sigma}$ vs bias voltage. The orange line displays Eq.~(3) from the Main text with $T = $\SI{11}{mK}.    
    \textbf{c}, Excess shot noise $S_{12}/(\tau_\mathrm{c}(1-\tau_\mathrm{c}))$ plotted versus the source noise 
    $S_\mathrm{\Sigma}$. Open symbols correspond to the measurement when the analyzer probes the inner channel. Full stars show $S_{12}/(\tau_\mathrm{c}(1-\tau_\mathrm{c})) - S_\mathrm{\Sigma}$ when the analyzer probes the outer channel. This data is the same as Supplementary Fig. \ref{Cross3tau}c, also for $\tau_\mathrm{s} = 0.5$. The two are expected to be the same. The continuous purple line displays the theory for $\delta t=$~\SI{64}{\pico\second}. Fits of the data in the region $V_\mathrm{s}>$~\SI{59}{\uV} (green dashed lines) give the Fano factors of $0.80$ (inner channel) and $0.73$ (outer channel).
	}
    \label{CrossInt}
\end{figure*}

In Supplementary Fig.~\ref{CrossInt} we show the cross-correlations when the charge is injected into the outer channels and the signal is measured in the inner channel. In that case, the positive contribution of the source noise is absent, and the signal is negative throughout. Like in the Main text, we show in Supplementary Fig.~\ref{CrossInt}a the transparencies of all quantum point contacts as function of bias, and in Supplementary Fig.~\ref{CrossInt}b that the source noise corresponds to the injection of charge $e$ (slope of the orange fit). We see a higher variation in the central QPC transparency, but this transparency is not expected to affect the signal (we will demonstrate this in Supplementary Fig. \ref{IndepTauC}). In Supplementary Fig.~\ref{CrossInt}c we show the cross-correlations of the inner channel, and the cross-correlations of the outer channel with the source noise subtracted. These are expected to coincide. They do show roughly the same slope (green lines), but there is some discrepancy between the curves themselves. We do not understand this discrepancy beyond earlier observations that the injection into the inner channel was not well controlled.  

\clearpage

\section{Cross-correlations in the full bias and temperature range}
\noindent

In Supplementary Fig.~\ref{ED_BlobAllV} we show the cross-correlations as function of source transparency in the full bias range and for the additional temperature of \SI{21}{mK}. As in Fig. \textbf{4} in the Main text the left/right column corresponds to the analyzer set to the outer/inner channel, respectively.

\begin{figure*}[!htb]
\renewcommand{\figurename}{\textbf{Supplementary Figure}}
\renewcommand{\thefigure}{\textbf{\arabic{figure}}}
	\centering
	\includegraphics[width=14.2cm]{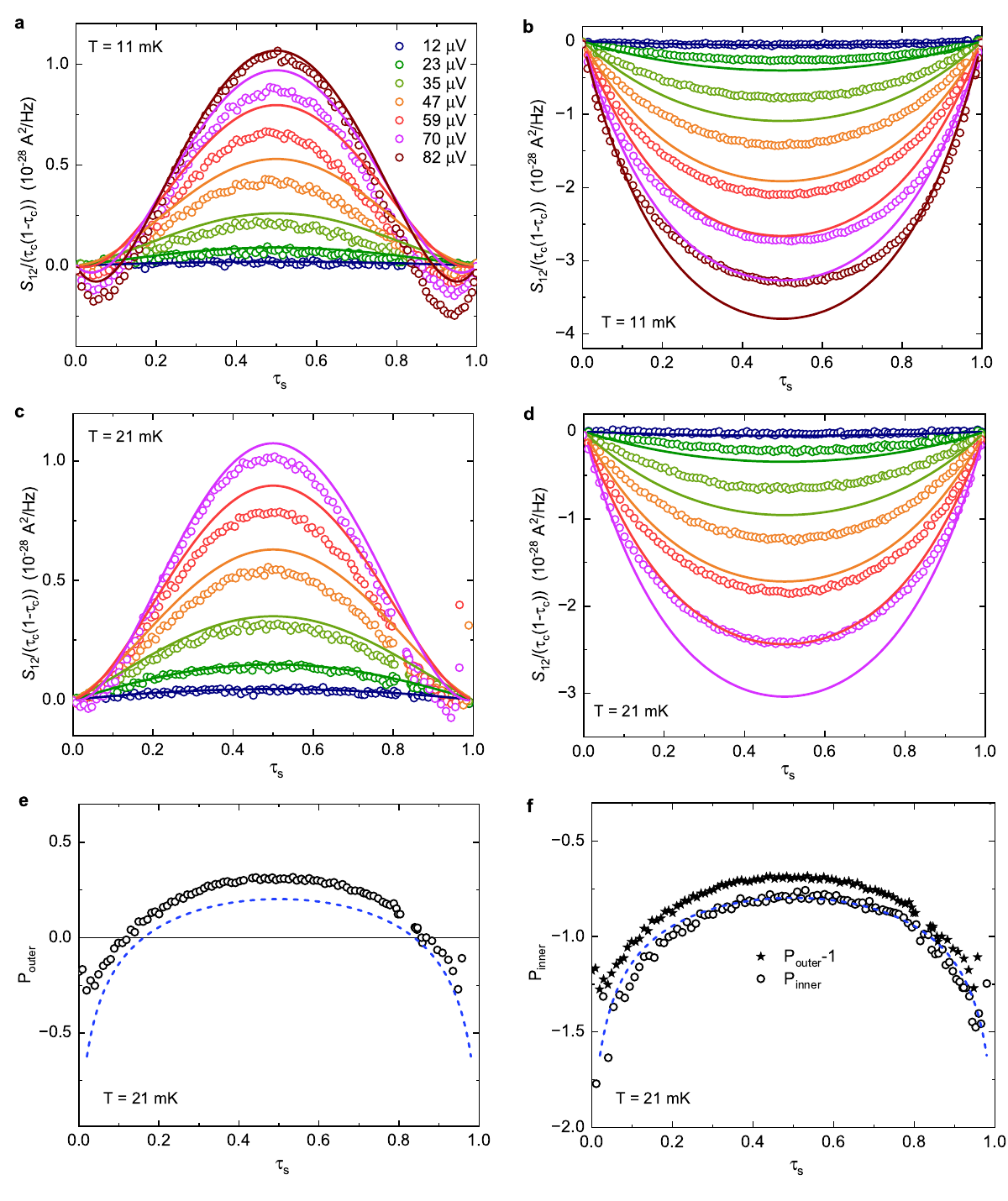}
	\caption{
	\footnotesize
	    \textbf{Cross-correlations vs dilution at intermediate voltages and higher temperature.}
	\textbf{a, c,} Cross-correlations $S_{12}/(\tau_\mathrm{c}(1-\tau_\mathrm{c}))$ as function of source QPC transmission $\tau_\mathrm{s}$ at \SI{11}{mK} (\textbf{a},\textbf{b}) and \SI{21}{mK} (\textbf{c}, \textbf{d}). Extension of Fig. \textbf{4} from the Main text. Injection takes place in the external channel at all times, whereas the measurement is done on the external channel (\textbf{a}, \textbf{c}, \textbf{e}) or the internal channel (\textbf{b}, \textbf{d}, \textbf{f}). The solid lines in \textbf{a}-\textbf{d} are the prediction for $\delta t =$ \SI{64}{ps}. Data at each bias voltage and their prediction have the same color (cf. legend). \textbf{e, f,} Fano factors extracted at $T=$ \SI{21}{mK} with the central QPC partially transmitting the outer (\textbf{e}) and inner (\textbf{f}) channel. Blue lines correspond to the high bias/large $\delta t$ prediction. Full stars in panel \textbf{f} display $P_\mathrm{outer}-1$. 
 }
 \label{ED_BlobAllV}
\end{figure*}


\section{Cross-correlations are independent of the analyzer transmission}
\noindent

\begin{figure*}[!htb]
\renewcommand{\figurename}{\textbf{Supplementary Figure}}
\renewcommand{\thefigure}{\textbf{\arabic{figure}}}
	\centering
	\includegraphics[width=10cm]{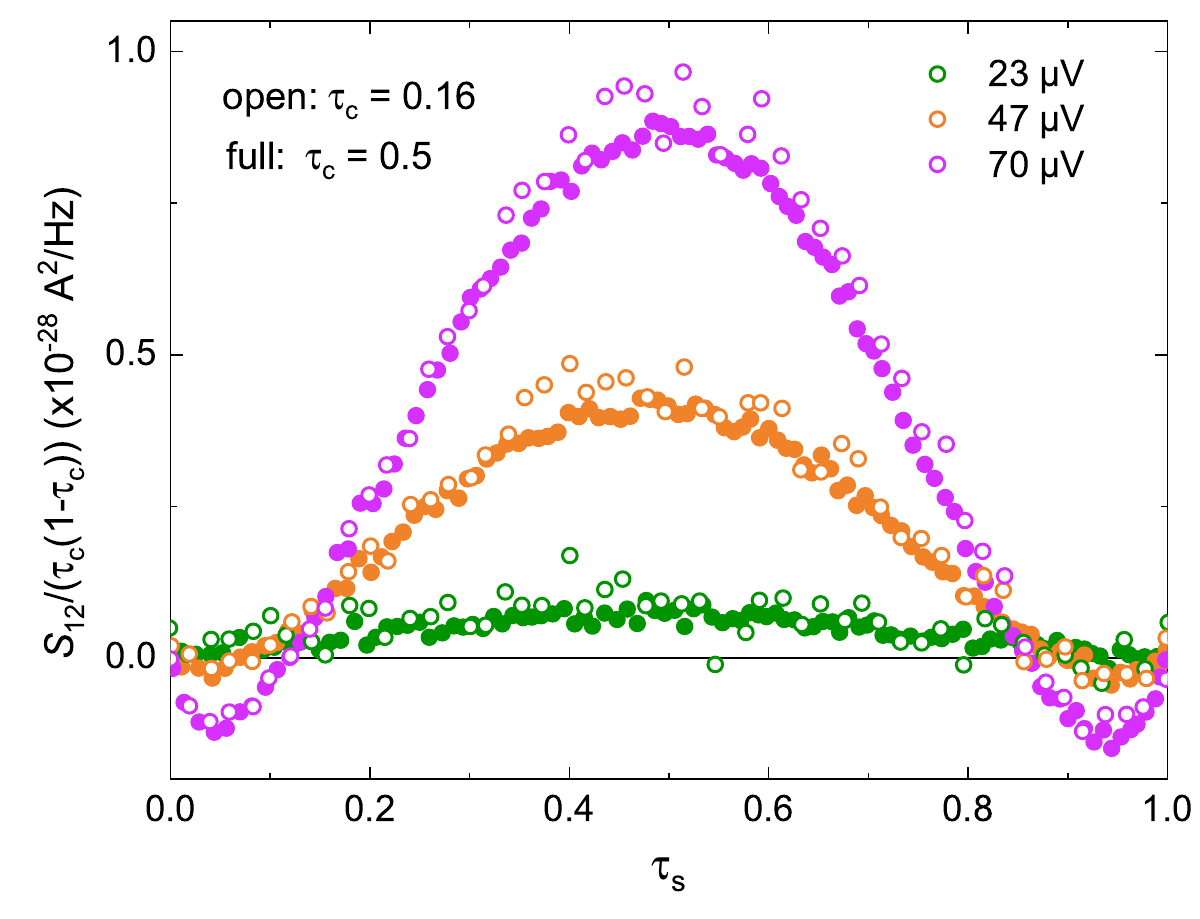}
	\caption{
	\footnotesize
	\textbf{}	
 Cross-correlations $S_{12}/(\tau_\mathrm{c}(1 - \tau_\mathrm{c}))$ as function of source QPC transmission $\tau_\mathrm{s}$ for $\tau_\mathrm{c}$ = 0.5 (open symbols) and $\tau_\mathrm{c}$ = 0.16 (full symbols).  Bias voltages are shown in the figure.
	}	\label{IndepTauC}
\end{figure*}

As mentioned, the cross-correlation signal $S_\mathrm{12}/(\tau_\mathrm{c}(1-\tau_\mathrm{c}))$ does not depend on the transparency of the analyzer. We have checked this by comparing the curves measured at $\tau_\mathrm{c} = 0.5$ and  $\tau_\mathrm{c} = 0.16$ at three bias voltages, see Supplementary Fig. \ref{IndepTauC}. As another control measurement, we have verified that there is no cross-correlation signal when the analyzer QPC is set to transmission $\tau_\mathrm{c}=1$, i.e., is on the plateau. This is expected as there is no partition at the analyzer QPC in that case.


\section{Tunneling between the inner and outer channel}
\noindent

\begin{figure*}[!htb]
\renewcommand{\figurename}{\textbf{Supplementary Figure}}
\renewcommand{\thefigure}{\textbf{\arabic{figure}}}
	\centering
	\includegraphics[width=16cm]{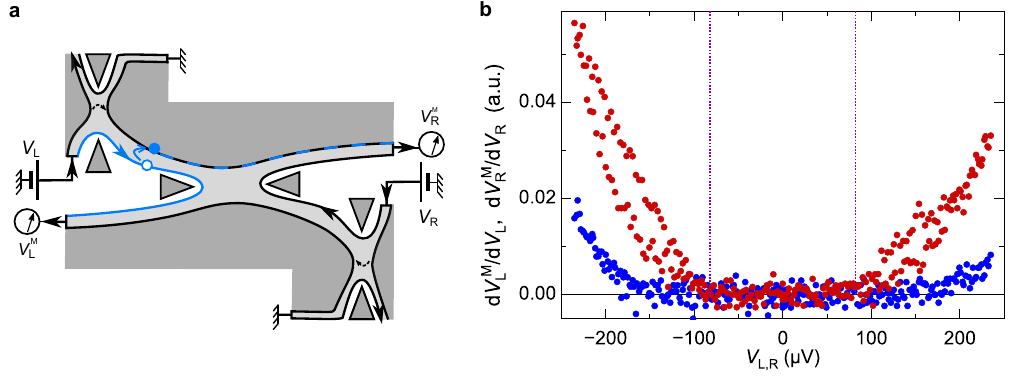}
	\caption{
	\footnotesize
	\textbf{a}, Schematics of the tunneling process and the measurement setup. \textbf{b}, Measured differential conductance due to tunneling as function of bias voltage. Blue and red dots correspond to the injection from the left/right side.
 }	\label{Tunneling}
\end{figure*}

At $\nu=2$, tunneling between the two adjacent copropagating channels is usually negligible. However, at long effective propagating distance, tunneling events can develop and alter our cross-correlation signal. Indeed, a carrier hopping from the outer channel to the inner one results in current fluctuation $\delta I$ on the inner channel and a correlated $-\delta I$ one in the outer channel. Therefore, such artifacts would create unwanted additional noise $-\tau_\mathrm{c}^2\delta I^2$ on the measured cross-correlations on the leads downstream to the central QPC.

We calibrate the tunneling by injecting energy on the outer channel while the central QPC is set on the plateau. In that configuration the only contribution to the signal is expected to come from tunneling. Therefore, measuring voltage $V_\mathrm{R(L)}^M$ at the frequency of $V_\mathrm{L(R)}$ can be directly attributed to the tunneling events along the path between the left (right) source and central QPC (see Supplementary Fig.~\ref{Tunneling}a). The fraction of current $dV_\mathrm{R(L)}^M/dV_\mathrm{L(R)}$ that tunnels between the edges is plotted in Supplementary Fig.~\ref{Tunneling}b as function of $V$. It is found to remain negligible in the  bias range used for the main measurements $V\leq$ \SI{82}{\uV}, indicated by purple vertical lines. Note that a hysteresis appears at higher voltage, prompting us to remain in the range $V \in [-82,82]$ \SI{}{\uV}.

\begin{figure*}[!htb]
\renewcommand{\figurename}{\textbf{Supplementary Figure}}
\renewcommand{\thefigure}{\textbf{\arabic{figure}}}
	\centering
	\includegraphics[width=16cm]{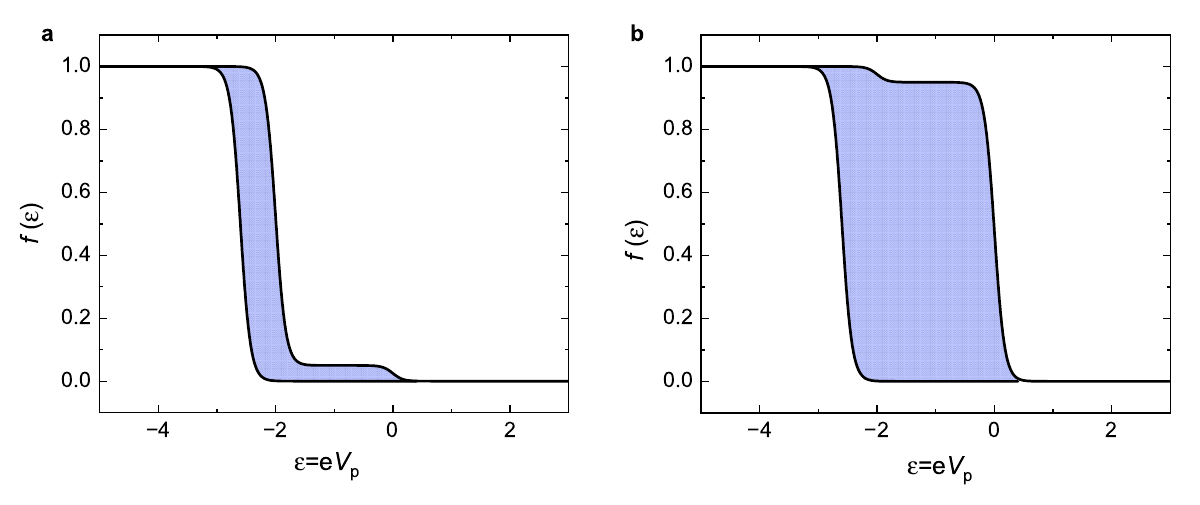}
	\caption{
	\footnotesize
	Assumed distribution functions between two adjacent edge channels after the injection of a quasiparticle into one of them. The available tunneling phase space from one to the other channel is shaded. Panel \textbf{a}/\textbf{b} refers to the source transparency $\tau_\mathrm{s} = 0.05/0.95$ respectively. 
 }	\label{Tunneling_dist}
\end{figure*}

Another effect of tunneling considered in the Main text is it being a potential cause of asymmetry in the measured cross-correlations at $\tau_\mathrm{s}=$ 0.05 and 0.95. 
If we consider the two distribution functions in the two copropagating edge channels after the injection of the quasiparticle into one channel, we get the situation shown in Supplementary Fig. \ref{Tunneling_dist} where the channel with injection is a double-step function, and the adjacent channel is a single-step function. The phase space for tunneling from one channel to the other is shaded. As we see, it is much larger for $\tau_\mathrm{s}=0.95$, resulting in extra negative cross-correlation signal, consistent with our observations.


\section{Oscillations}
\noindent

The non-perturbative theory prediction shows oscillations with $(\delta t)^{-1}$ which are not found in the measurement. We remind that $\delta t=d/v_\mathrm{n}-d/v_\mathrm{c}$ is the time delay between fractionalized wave-packets, with $d$ the distance between source and analyzer quantum point contacts and $v_\mathrm{c,n}$ the velocities of charged and neutral mode. Moreover, $\delta t$ is the only fitting parameter. In Supplementary Fig. \ref{Oscillations} we numerically go to much higher bias than available experimentally in order to explore the asymptotic behavior. We expect the oscillations to dampen with bias and at sufficiently high bias to not have a difference between finite and infinite $\delta t$.

This is indeed what we find. The finite $\delta t$ curves (black) oscillate above (low transmission) or under (high transmission) the corresponding $\delta t=\infty$ curves (blue), and, at high enough bias the black and blue curve coincide. Since the slope is the Fano factor (up to a multiplicative constant), we conclude that it should be calculated at $\delta t = \infty$. 

\begin{figure*}[!htb]
\renewcommand{\figurename}{\textbf{Supplementary Figure}}
\renewcommand{\thefigure}{\textbf{\arabic{figure}}}
	\centering
	\includegraphics[width=16cm]{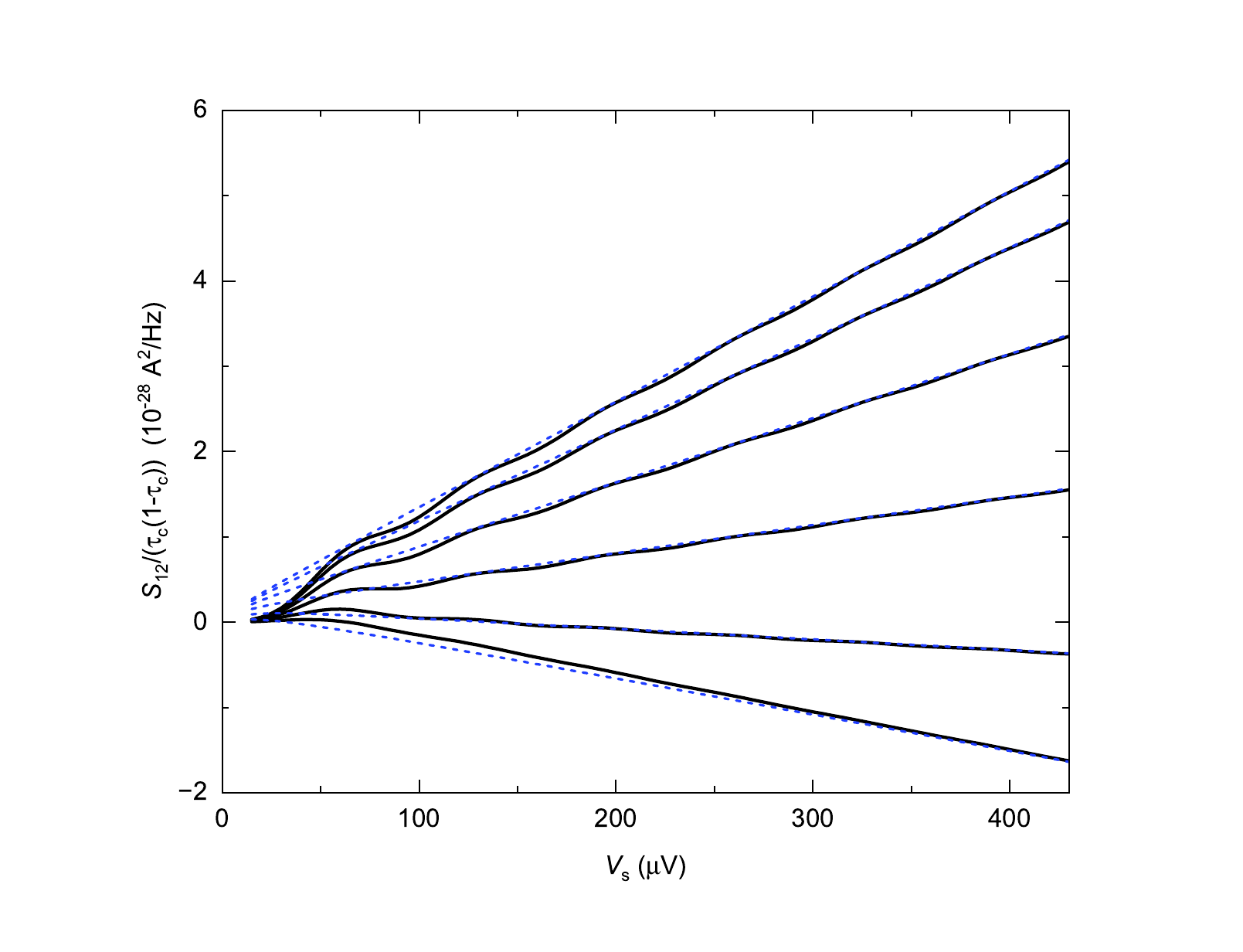}
	\caption{
	\footnotesize
	Comparison of calculated cross-correlations for $\delta t$ = \SI{64}{ps} (black curves) and $\infty$ (blue curves). Transmissions $\tau_\mathrm{s}$ range from 0.06 to 0.46 in steps of 0.08 (bottom to top). 
 }	\label{Oscillations}
\end{figure*}

We assume that the oscillations are due to cutoffs at finite energies $k_\mathrm{B}T$ and $h/\delta t$. 
The discrepancy between data and theory in the collider geometry (as opposed to distributions) may be at least partially due to these oscillations. 

\vspace{30mm}

\section{Limitations of the non-perturbative model}
\noindent

We note that some experimental details are beyond the scope of the model. 
The model 
does not include long-range interactions, dissipation, plasmon dispersion, or the coupling of the plasmon modes to the adjacent charge puddles. 




\bibliographystyle{naturemag}

\end{document}